\documentclass[fleqn,usenatbib]{mnras_tex_edited}

\usepackage{amssymb,amsmath,verbatim,mathtools,needspace,enumitem,etoolbox,graphicx,physics,microtype,natbib,url,hyperref,mathrsfs,bm,ulem,lipsum,xspace}
\usepackage{orcidlink}

\renewcommand{\emph}[1]{\textit{#1}}
\newcommand{\authorlinebreak}{\medskip \smallskip \\ \noindent \LARGE \rm  $\qquad\!\!\!\!\!\!\!\!\!\!\!\!\!\!\!$
}
\newcommand{\approptoinn}[2]{\mathrel{\vcenter{
  \offinterlineskip\halign{\hfil$##$\cr
    #1\propto\cr\noalign{\kern2pt}#1\sim\cr\noalign{\kern-2pt}}}}}

\defcitealias{2003MNRAS.341..501S}{SG03}
\defcitealias{2005ApJ...630..167T}{TQM05}
\newcommand{\SK}{{\citetalias{2003MNRAS.341..501S}}\xspace}
\newcommand{\SKO}{{\cite{2003MNRAS.341..501S}}\xspace}
\newcommand{\TP}{{\citetalias{2005ApJ...630..167T}}\xspace}
\newcommand{\TPO}{\cite{2005ApJ...630..167T}\xspace}
\newcommand{\Teff}{T_{\rm eff}}
\newcommand{\tauv}{\tau_{\rm v}}
\newcommand{\cs}{c_{\rm s}}
\newcommand{\Mdot}{\dot{M}}

\newcommand{\kb}{k_{\rm B}}
\newcommand{\Rout}{r_{\rm out}}
\newcommand{\Mdotout}{\dot{M}_{\rm out}}

\newcommand{\der}{\mathrm{d}}

\newcommand{\codename}{\texttt{pAGN}\xspace}

\newcommand{\milan}{Dipartimento di Fisica ``G. Occhialini'', Universit\'a degli Studi di Milano-Bicocca, Piazza della Scienza 3, 20126 Milano, Italy}
\newcommand{\infn}{INFN, Sezione di Milano-Bicocca, Piazza della Scienza 3, 20126 Milano, Italy}
\newcommand{\bham}{School of Physics and Astronomy \&	 Institute for Gravitational Wave Astronomy, University of Birmingham,\vspace{-0.05cm}\\$\;$
Birmingham, B15 2TT, UK}

\title[AGN disc modeling]{\codename:
the one-stop solution for AGN disc modeling
}

\author[Gangardt et al.]{Daria Gangardt$\,$\orcidlink{0000-0001-7747-689X}$^1$, 
Alessandro Alberto Trani$\,$\orcidlink{0000-0001-5371-3432}$^{2,3,4}$, 
Clément Bonnerot$\,$\orcidlink{0000-0001-9970-2843}$^1$, 
\authorlinebreak
Davide Gerosa$\,$\orcidlink{0000-0002-0933-3579}$^{5,6,1}$
\bigskip
\\
$^{1}$\bham\\
$^{2}$Niels Bohr International Academy, Niels Bohr Institute, Blegdamsvej~17, Copenhagen, 2100, Denmark\\
$^{3}$Research Center for the Early Universe, The University of Tokyo, 7~Chome-3 Hongo, Bunkyo-ku, 113-0033, Tokyo, Japan\\
$^{4}$Okinawa Institute of Science and Technology, 1919-1 Tancha, Onna-son, 904-0495, Okinawa, Japan\\
$^{5}$\milan\\
$^{6}$\infn
}

\begin{document}
\label{firstpage}
\pagerange{\pageref{firstpage}--\pageref{lastpage}}
\maketitle

\begin{abstract}

Models of accretion discs surrounding active galactic nuclei (AGNs) find vast applications in %
 high-energy astrophysics. The broad strategy is to parametrize %
  some of the key disc properties such as gas density and temperature as a function of the radial coordinate from a given set of assumptions on the underlying physics. Two of the most popular approaches in this context were presented by Sirko \& Goodman (2003) and Thompson et al. (2005). We present a critical reanalysis of these widely used models, detailing their assumptions and clarifying some steps in their derivation that were previously left unsaid. Our findings are implemented in the \codename module for the Python programming language, which is the first public implementation of these accretion-disc models. We further apply \codename to the evolution of stellar-mass black holes embedded in AGN discs, addressing the potential occurrence of migration traps.
\end{abstract}


\section{Introduction}
\label{sec:Intro}

Active galactic nuclei (AGNs) are compact regions at the center of galaxies powered by gas accretion onto supermassive black holes (BHs) as opposed to solely the radiation from stars. The underlying theory describing the accretion disc of AGNs was first introduced by \cite{1964SPhD....9..195Z} and \cite{1964ApJ...140..796S}. 
AGNs have been studied at low and high redshifts across several electromagnetic wavelengths, capturing a wide range of astrophysical phenomena (see \citealt{2015ARA&A..53..365N, 2017A&ARv..25....2P, 2018ARA&A..56..625H, 2022hxga.book....4B} for broad reviews on the topic).
Due to the deep gravitational well surrounding the central BH, the gas in the accretion disc is expected to reach temperatures of ${\sim} 10^5$K and surface densities of ${\sim} 10^{5}$g cm$^{-2}$.
The accretion disc of the central BH extends to sub-pc scales and is surrounded by optically thick material which is coupled to the disc itself. %
These components are 
collectively referred to as the AGN disc, which is expected to extend
to separations of 1-10 pc \citep{2015ARA&A..53..365N}.
Because of high obscurations and uncertainty in observations, the actual size of AGN discs is somewhat unclear but tends to be larger than what is expected from theoretical models 
\citep{2022MNRAS.511.3005J,2022ApJ...929...19G,2022ApJ...940...20G}.

AGN discs are unique astrophysical environments with a rich phenomenology, including high-energy jets, dusty torii, and accreting BHs \citep{2017A&ARv..25....2P}.
In the context of gravitational-wave observations, AGN discs are studied as host environments for compact-binary formation and mergers \citep{2012MNRAS.425..460M, 2011MNRAS.417L.103M, 2019PhRvL.123r1101Y, 2019ApJ...878...85S, 2020MNRAS.499.2608F, 2020ApJ...898...25T, 2024A&A...683A.135T}. The large escape velocity around a supermassive BH implies that objects are likely to be retained in the disc vicinities,  potentially forming a large population of stellar-mass BHs that have a higher likelihood of interacting. The dense gas in the disc can facilitate binary formation, accelerate the inspiral, and induce chains of hierarchical BH mergers \citep{2021NatAs...5..749G, 2023PhRvD.108h3033S, 2023arXiv230911561W, 2023arXiv231118548V}. The occurrence of hierarchical mergers in AGN discs crucially depends on the presence of the so-called migration traps, namely locations in the disc where the migration torque changes sign, which is still an open issue in AGN-disc modeling \citep{2016ApJ...819L..17B, 2020ApJ...898...25T, 2024MNRAS.530.2114G}. %

Early models of AGNs discs consist of one-dimensional, steady-state, semi-analytic solutions utilizing parametric prescriptions.  %
Subsequent computational advancements %
allowed for models capturing more complex physics, such as radiative transfer, gas phase transitions, magnetic fields, and general relativity \citep[e.g.][]{1972ApJ...173..431W,1984ApJ...277..296H,1991MNRAS.250..581F,2005A&A...437..861S,2008A&A...482...67S,2009ApJ...702...63W,2023MNRAS.520.5090H}.
Nevertheless, one-dimensional models remain highly valuable today due to their computational efficiency and insightful perspectives on the structure of AGN discs. This makes them particularly useful in the study of interactions between compact objects and BHs.
The first of these one-dimensional approaches dates back to \cite{1973A&A....24..337S}, who first model geometrically thin, optically thick discs around a BHs.  Building on this seminal work, two models are most commonly used in the field, namely those by \cite{2003MNRAS.341..501S} and \cite{2005ApJ...630..167T} (but see also \citealt{2003MNRAS.339..937G, 2003astro.ph..7084L, 2009ApJ...700.1952H, 2021ApJ...910...94C, 2022ApJ...928..191G, 2024MNRAS.530.2114G,2024OJAp....7E..19H, 2024OJAp....7E..20H}).
Both these models assume some heating mechanisms in the disc that marginally support the outer regions from collapsing due to self-gravity and formulate one-dimensional sets of equations for the AGN-disc profile as a function of a number of parameters such as the  mass of the central BH and  the accretion rate. 

The \citet{2003MNRAS.341..501S} and \cite{2005ApJ...630..167T} models are widely used and underpin some of the key, qualitative results in the field of AGN-disc physics. Despite that, the underlying parameters and methods are often left unspecified.
Achieving a stable numerical implementation of these disc solutions is not straightforward and codes in this area have not been released in the public domain. The goal of this paper is to critically re-analyze the AGN disc models by \citet{2003MNRAS.341..501S} and \citet{2005ApJ...630..167T}.  In particular, we clarify the model equations one needs to solve (and crucially the order one needs to solve them), highlight the choices one has to make to obtain stable solutions, and provide a  highly customizable implementation. %
Our software is made publicly available in a Python package called \codename (short for ``parametric AGNs", pronounced as ``pagan'').

This paper is organized as follows.
In Sec.~\ref{sec:AGNeqs}, we lay out the equations for the \citet{2003MNRAS.341..501S} and \citet{2005ApJ...630..167T} models. 
In Sec.~\ref{sec:disc}, we explore some of the input parameter space for both models. In Sec.~\ref{sec:migration}, we showcase our implementation, looking in particular at the occurrence of migration traps in either of the two disc models. In Sec.~\ref{sec:code}, we present the public code \codename. In Sec.~\ref{sec:concl}, we draw our conclusions and present prospects for future work.

\begin{table}
\centering
  \caption{Key parameters entering the \citet{2003MNRAS.341..501S} and \citet{2005ApJ...630..167T} AGN-disc models. The third column indicates whether the parameter is an input of the model (I), a fixed value for the entire disc (F), or a profile parameter obtained by running the model (P). The accretion rate $\Mdot$ is a fixed parameter for the \SK disc but a profile parameter for the \TP disc.
  \label{tab:params} }
\begin{tabular}[c]{| l | c | c |}
\hline
 \multicolumn{1}{|c|}{\textbf{Symbol}} & \multicolumn{1}{c|}{\textbf{Definition}} & \multicolumn{1}{|c|}{\textbf{I/F/P}} \\ \hline
 $M$ & Mass of the central BH & I\\
  $R_\mathrm{s}$ & BH Schwarzschild radius& F\\
  $L_\mathrm{Edd}$ & Eddington luminosity & F \\
  $\Mdot_\mathrm{Edd}$ & Eddington accretion rate& F\\
  $X$ & Hydrogen abundance in disc & I\\
          $\kappa_{\rm es}$ & Electron scattering opacity & F \\
    $r$ & Radial distance from the central BH& P \\
      $\Mdot$ & Mass accretion rate& F or P \\
  $r_{\rm min}$ & Inner edge of the disc & I \\
    $T$ & Midplane temperature& P\\
     $\Teff$ & Midplane effective temperature& P \\
      $\rho$ & Midplane density& P\\
       $h$ & Height of disc from the midplane & P\\
       $\Sigma_\mathrm{g}$ & Midplane surface density& P \\
       $\Sigma_\mathrm{tot}$ & Midplane total dynamical density& P\\
       $f_{\rm g}$ & Gas fraction& P\\
       $\tauv$ & Midplane optical depth& P \\
       $\kappa$ & Midplane opacity& P\\
        $c_\mathrm{s}$ & Midpane sound speed & P\\
         $p_{\rm gas}$ & Gas pressure & P\\
          $p_{\rm rad}$ & Radiation pressure & P\\
 \hline\hline \\[-8pt]
\multicolumn{3}{|c|}{\citet{2003MNRAS.341..501S} parameters} \\
\hline
 \multicolumn{1}{|c|}{\textbf{Symbol}} & \multicolumn{1}{c|}{\textbf{Definition}}  & \multicolumn{1}{c|}{\textbf{I/F/P}} \\ \hline
 $\alpha$ & Shakura-Sunyaev viscosity parameter & I\\
 $r_{\rm max}$ & Outer edge of the disc & I\\
 $l_\mathrm{E}$& Disc Eddington ratio & I \\
    $\epsilon_\mathrm{S}$ & Radiative efficiency& I\\
     $b$ & Switch for viscosity-pressure relation & I\\
     $Q_\mathrm{S}$ & Toomre stability parameter & P\\
  $\nu$ & Disc viscosity & P \\
     $\Omega_\mathrm{S}$ & Rotational velocity & P\\
     $\beta$ & Gas pressure to total pressure ratio & P\\
  \hline \hline \\[-8pt]
\multicolumn{3}{|c|}{\citet{2005ApJ...630..167T} parameters} \\
 \hline
 \multicolumn{1}{|c|}{\textbf{Symbol}} & \multicolumn{1}{c|}{\textbf{Definition}}  & \multicolumn{1}{c|}{\textbf{I/F/P}} \\ \hline
 $\sigma$ & Stellar dispersion velocity& I\\
 $\Rout$ & Effective outer edge of the disc & I\\
  $\Mdotout$ & Accretion rate at $\Rout$ & I\\
   $m_\mathrm{T}$ & Global torque efficiency & I\\
    $\epsilon_\mathrm{T}$ & Star formation efficiency & I\\
     $\xi$ & Supernova radiative efficiency & I\\
     $\dot{\Sigma}_\mathrm{*}$ & Star formation rate & P \\
$\eta$ & Star formation efficiency fraction & P\\
 $Q_\mathrm{T}$ & Toomre stability parameter & P\\
 $\Omega_\mathrm{T}$ & Rotational velocity& P \\
\hline \hline

 \end{tabular}	
 
 \end{table}

\begin{figure*}
    \centering
    \includegraphics[scale=0.76]{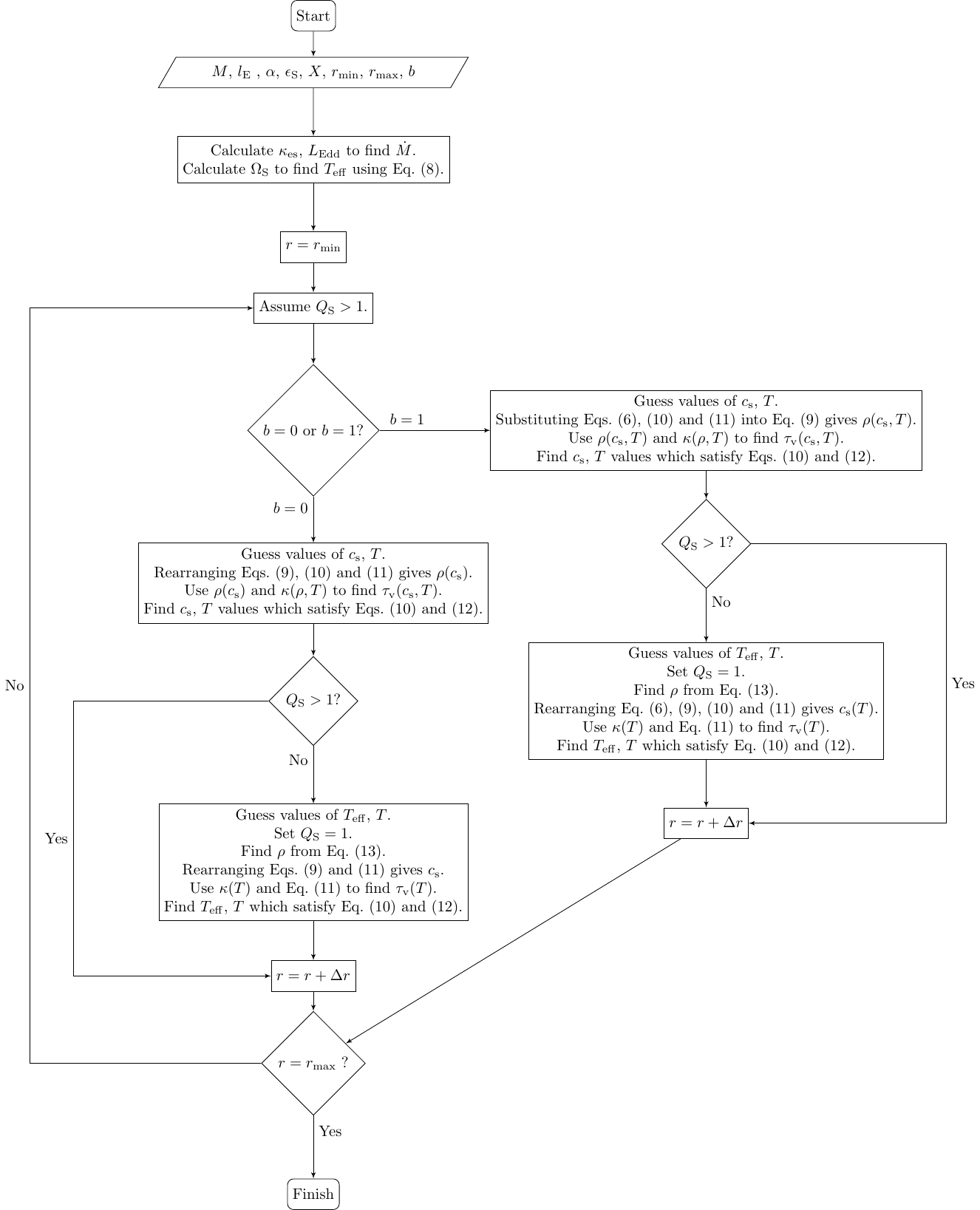}
    \\[15pt]
    \caption{Flowchart showing detailing our solution strategy for the \SK model. Construction proceeds from the inner disc to the outer disc, with initial guesses on the stability parameter $Q_{\rm Q}$ which are then checked a-posteriori. 
    }\label{fig:sirko_flowchart}
\end{figure*}

\begin{figure*}
    \centering
    \includegraphics[scale=0.76]{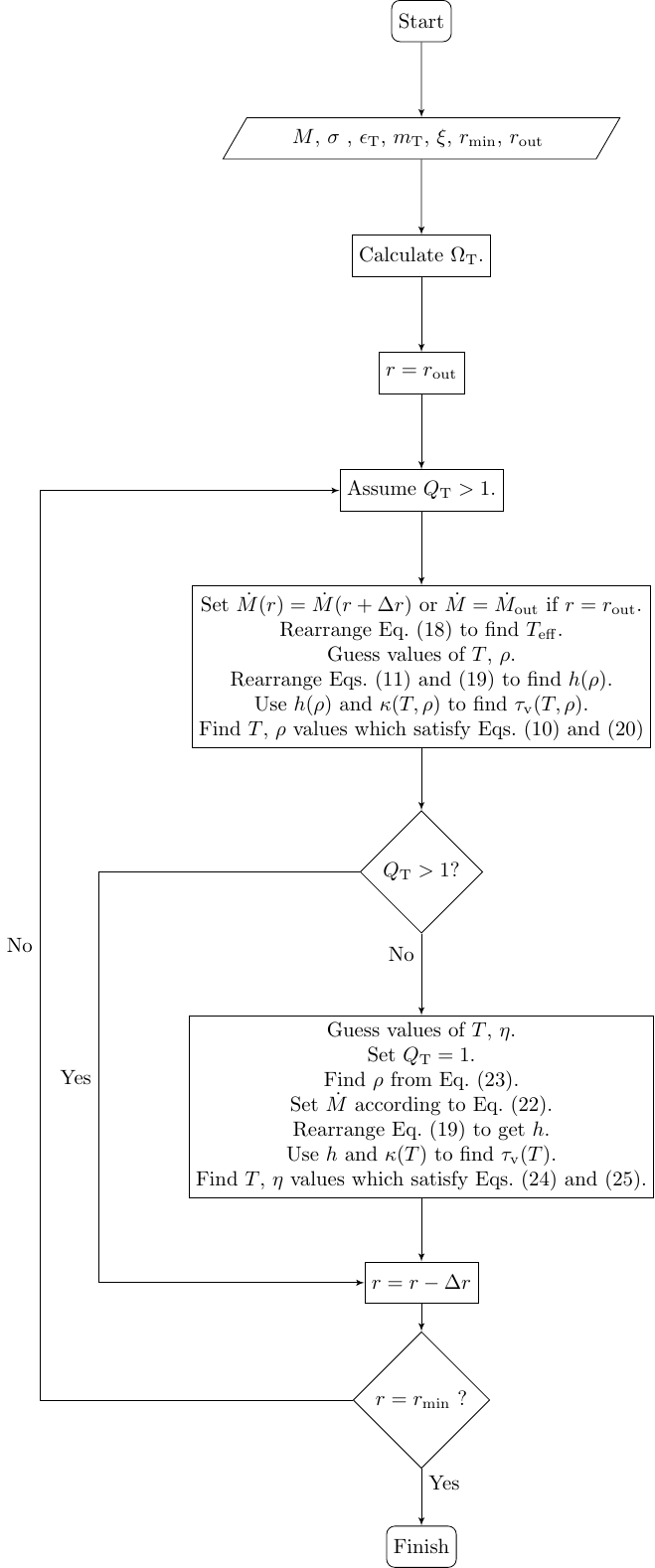}
    \\[15pt]
    \caption{Flowchart showing detailing our solution strategy for the \TP model. Construction proceeds from the outer disc to the inner disc, with initial guesses on the stability parameter $Q_{\rm T}$ which are then checked a-posteriori.
    }
    \label{fig:thompson_flowchart}
\end{figure*}

\section{AGN disc Models } \label{sec:AGNeqs}
We first summarize the AGN disc models by \cite{2003MNRAS.341..501S} and 
\cite{2005ApJ...630..167T}. 
We refer to the models as \SK and \TP, respectively. Both models consist of an inner, thin accretion disc extended to larger radii to explain observed AGN luminosities. %
In the outer regions, the disc needs to remain marginally stable against fragmentation. %
With respect to the \citet{1973A&A....24..337S} thin-disc solution, the \SK model additionally assumes the existence of some heating mechanism generating radiation pressure that can support the outer parts of the disc against collapse. The \TP model further modifies the \SK model, with the most notable change being that the mass advection is driven by non-local torques rather than local viscous stresses. 
Furthermore, the \SKO accretion rate is constant across the disc while that of \TPO varies because it directly takes into account the mass lost to star formation. 

We now introduce each model in closer detail and present the key equations one needs to solve to build the resulting disc profiles. For clarity, the parameters entering each model are reported in Table \ref{tab:params}. A step-by-step guide on how the equations are solved is provided in Figs.~\ref{fig:sirko_flowchart} and~\ref{fig:thompson_flowchart}.

\subsection{Sirko \& Goodman (2003)}

\subsubsection{Modeling strategy}

In the inner regions, the \SK{} model assumes a thin and viscous accretion disc %
to be the source of AGN luminosity (as proposed by \citet{1981ARA&A..19..137P}), similar to the disc model by \citet{1973A&A....24..337S}. Such a self-gravitating disc cannot be extended to large radii, where the gravitational pull in the vertical direction causes disc fragmentation and star formation, thus depleting the disc of gas to sufficiently fuel the inner regions. %
The \SK model resolves this by assuming that some auxiliary heating (i.e. heating that does not come from orbital energy) lowers the density of the gas in the outer region gas, thus reducing the gravitational pressure. This heating is most likely sourced by star formation, but this is left unspecified in the \SK model. 
The auxiliary heating process is prescribed so that gas supply from the marginally gravitationally stable outer regions keeps fueling the hotter inner regions all under a constant gas accretion rate $\dot{M}$.

The stability of the disc is encoded by the parameter first defined by \citet{1964ApJ...139.1217T} for circular Keplerian orbits
\begin{equation} \label{eq:SirkoQ}
    Q_\mathrm{S} \equiv \frac{c_\mathrm{s} \Omega_\mathrm{S}}{\pi G \Sigma_\mathrm{g}} \approx \frac{\Omega_\mathrm{S}^2}{2 \pi G \rho} \, ,
\end{equation}
 where $c_\mathrm{s}$ is the speed of sound, $\Omega_\mathrm{S} = \sqrt{GM/r^3} $ is the angular velocity of the disc, $\Sigma_\mathrm{g} = 2 \rho h$ is the midplane mass surface density, $\rho$ is the midplane mass density, and $h$ is the height from the midplane. 
The disc collapses and fragments whenever $Q_\mathrm{S}<1$.  The \SK model is made of two regimes. In the inner region one has $Q_\mathrm{S} \gg 1$: the angular frequency and temperature are high and there is no risk of fragmentation. The outer region instead presents $Q_\mathrm{S} \sim 1$: the disc is only marginally stable and auxiliary heating sources become necessary to prevent vertical collapse.

The construction of the model proceeds from an inner boundary $r_{\rm min}$ and assumes a zero-torque boundary condition, see Fig.~\ref{fig:sirko_flowchart}.  \SKO approximate the innermost stable circular orbit to be $r_{\rm min} = R_\mathrm{s} / 4 \epsilon_{\rm S}$, where $R_\mathrm{s} = 2 GM /c^2$ is the Schwarzschild radius of the BH and $\epsilon_\mathrm{S}$ is the radiative efficiency of the BH, which is set to $\epsilon_\mathrm{S} = 0.1$.
For each gas ring at a cylindrical radius $r$ from the central BH, 
one first assumes that the ring is located in the inner regime where $Q_\mathrm{S} \gg 1$. The equations presented in Sec.~\ref{sec:Sirkoinner} below are then solved to find $\Omega_{\rm S}$ and $\rho$. In turn these are used to evaluate $Q_\mathrm{S}$  from Eq.~(\ref{eq:SirkoQ}). If $Q_{\rm S}<1$, one switches to the $Q_\mathrm{S}=1$ regime and solves the equations from Sec. \ref{sec:Sirkoouter} instead. This process is then repeated for every value of $r$ until $r = r_{\rm max}$.
Unless specified, we set $r_{\rm max}$ 
 to the minimum between $10^7 R_{\rm s}$ and $1$~pc. An unreasonably large value of $r_{\rm max}$ %
leads to a spectral energy distribution that does not match observations (cf. \citealt{2003MNRAS.341..501S}). %

The accretion rate of the \SK disc is parameterized by the Eddington ratio 
\begin{equation}
l_{\rm E}  =  \frac{\Mdot \epsilon_{\rm S} c^2 }{ L_{\rm Edd}}\,,
\end{equation}
where $L_{\rm Edd}$ is the Eddington luminosity and the normalization is set to the luminosity %
of a non-self gravitating disc. In turn, the Eddington luminosity is
\begin{equation}\label{eq:Ledd}
    L_{\rm Edd} = \frac{4 \pi G M c}{\kappa_{\rm es}} \,,
\end{equation}
where $\kappa_{\rm es} = 0.2(1 + X) \, {\rm cm}^2 {\rm g}^{-1}$ 
is the electron scattering opacity for a fractional abundance of hydrogen which we assume to be $X=0.7$. The \SK model thus depends on the mass of the central BH $M$ through both the angular velocity of the disc $\Omega_{\rm S}$ and the accretion rate $\Mdot$. %

The disc viscosity is prescribed using the  \cite{1973A&A....24..337S} dimensionless parameter
\begin{equation}
\alpha = \frac{\nu}{ c_{\rm s} h \beta^b} \,,
\end{equation}
where 
\begin{equation}
\beta = \frac{p_{\rm gas}}{p_{\rm gas} +p_{\rm rad}} = \frac{p_{\rm gas}}{p_{\rm tot}}
\end{equation}
is the fraction of gas pressure $p_{\rm gas}$ to total pressure $p_{\rm tot}$; the latter contains contributions from both gas and radiation. The parameter $b=\{0,1\}$  acts as a switch flag to determine how viscosity and pressure relate in the disc. The two cases are often referred to as $\alpha$-disc ($b = 0$) and $\beta$-disc ($b=1$), see e.g. \cite{2009ApJ...700.1952H}. 
For the gas pressure, we use the ideal gas law 
\begin{equation}\label{eq:gasp}
p_{\rm gas} = \frac{\rho \kb T }{ m_\mathrm{U}}
\end{equation} 
where $\kb$ is the Boltzmann constant and  $m_\mathrm{U}$ is the atomic-mass constant. 
The radiation pressure is given by
\begin{equation} 
p_{\rm rad} = \frac{ \sigma_{\rm SB} \tauv}{2 c}T_{\rm eff}^4\,,
\end{equation}
 which is constructed such that in the optically thick regime it recovers $p_{\rm rad} = 4\sigma_{\rm SB}T^4/3c$, but retains a dependency on $\tauv$ in the optically thin regime \citep{2003MNRAS.341..501S}. The source of the radiation pressure is not made explicit by \SKO, but is assumed to come from stellar processes such as supernovae and nuclear fusion in stars.

\subsubsection{Inner regime} \label{sec:Sirkoinner}

For each value of $r$, the model first assumes that there is no star formation ($Q_\mathrm{S} \geq 1$). Each annulus is treated as a black-body with an effective temperature $\Teff$. This is found by equating the locally radiated flux to the viscous heating rate per unit area \citep{1973A&A....24..337S}:
\begin{equation}\label{eq:SirkoTeffMdot}
    \sigma_{\rm SB} \Teff^4 = \frac{3 \Omega_\mathrm{S}^2}{8 \pi} \Mdot \left(1 - \sqrt{\frac{r_{\rm min}}{r}} \right) = \frac{3\Omega_\mathrm{S}^2}{8\pi} \Mdot' \, ,
\end{equation}
where $\sigma_{\rm SB}$ is the Stefan-Boltzmann constant and we have defined $\Mdot' = \Mdot (1 - \sqrt{{r_{\rm min}}/{r}} )$. %
Equation (\ref{eq:SirkoTeffMdot}) assumes that all material below $r = r_\mathrm{min}$ falls into the BH and cannot energetically interact with the rest of the disc.

Mass conservation relates the viscosity of the gas ring to the accretion rate \citep{2003MNRAS.341..501S}
 \begin{equation} \label{eq:Sirkoviscosityaccretion}
    \beta ^b \cs^2 \Sigma_{\rm g} = \frac{\Mdot' \Omega_\mathrm{S} }{3 \pi \alpha} \, ,
\end{equation}
which gives two families of solutions: $b = 0$ (where the viscosity is proportional to total pressure) and  $b = 1$ (where the viscosity is proportional to the gas pressure only).
The sound speed in the disc is defined as 
\begin{equation} \label{eq:Sirkocs}
    c_\mathrm{s}^2 = \frac{p_{\rm tot}}{\rho} \, .
\end{equation}
In this regime, for each value of $r$, we look for solutions in $c_{\rm s}$ and $T$ and rearrange all other parameters as functions of $c_{\rm s}$ and $T$ only. The midplane height can be expressed as a function of $c_{\rm s}$ by assuming hydrostatic equilibrium
 \begin{equation}\label{eq:Sirkoh}
 h = \frac{c_\mathrm{s} }{ \Omega_\mathrm{S}} \, .
 \end{equation}
  The value of the density as a function of $c_{\rm s}$ and $T$ is then given by substituting $\Sigma_\mathrm{g} = 2 \rho h$, Eq.~(\ref{eq:Sirkocs}) and Eq.~(\ref{eq:Sirkoh})  into 
Eq.~(\ref{eq:Sirkoviscosityaccretion}) for the $b=0$ case, and combining them with the equation for the gas pressure [see Eq.~(\ref{eq:gasp})] for the $b=1$ case.

The temperature profile in the disc depends on the optical depth $\tauv = \kappa \rho h$, where $\kappa (\rho, T)$ is the opacity. The latter is obtained using interpolated values by \cite{2003A&A...410..611S} when $T < 10^4 
{\rm K}$ and \cite{2005MNRAS.360..458B} when $T > 10^4 
{\rm K}$, the set of which we refer to as the ``combined'' opacity. These are newer prescriptions for the opacity compared to those by \cite{1996ApJ...464..943I} and \cite{1994ApJ...437..879A} used by \SKO. The opacities in \cite{2003A&A...410..611S} are calculated for silicate grains; the effect of graphite that is important at temperatures of ${\sim} 2000$ K and may be responsible for the broad line region in AGN observations \citep[see][]{2018MNRAS.474.1970B} is ignored. The inclusion of the effect of graphite in \codename is left to future work. From the opacity and effective temperature, we look for solutions in $T$ by assuming the disc ring is in radiative equilibrium:
\begin{align}\label{eq:Sirkoradiationequilibrium}
    T^4 = \left( \frac{3}{8}\tauv + \frac{1}{2} + \frac{1}{4\tauv}\right) \Teff^4 \, ,
\end{align}
where the functional form of the equation was chosen to match the temperature dependence on $\Teff$ and $\tauv$ across both the optically-thick and optically-thin regimes, cf.  \SKO. Finally, one can look for solutions in $c_\mathrm{s}$ and $T$ by considering Eq.~(\ref{eq:Sirkocs}).

The Toomre stability parameter $Q_\mathrm{S}$ is calculated from the second expression in Eq.~(\ref{eq:SirkoQ}). If this falls below $1$, it is assumed that the ring is in the outer regime and a different set of equations is used, which we present next.

\subsubsection{Outer regime} \label{sec:Sirkoouter}

In the outer regions, the model expects the disc to be only marginally gravitationally stable, i.e. $Q_\mathrm{S}=1$. In this case, Eq.~(\ref{eq:SirkoTeffMdot}) no longer applies since there is additional auxiliary heating. The density is then given by
\begin{equation} \label{eq:SirkoQrho}
    \rho = \frac{\Omega_\mathrm{S}^2}{2 \pi G},
\end{equation}
which is a rearrangement of Eq.~(\ref{eq:SirkoQ}) where $ Q_\mathrm{S} = 1$. 
In the inner regime we look for solutions in $c_{\rm s}$ and $T$; in the outer regime we know the value of the density $\rho$ and instead look for solutions in $T$ and $\Teff$. This is done by rearranging Eq.~(\ref{eq:Sirkoviscosityaccretion}) and substituting in Eqs.~(\ref{eq:Sirkocs}), (\ref{eq:Sirkoh}) and (\ref{eq:SirkoQrho}) to obtain an expression for $c_{\rm s}$ as a function of $T$. In the $b=0$ case, $c_\mathrm{s}$ can be independently determined for a given value of $r$; in the $b=1$ case, $c_{\rm s}$ is a function of the temperature $T$ [see Eq.~(\ref{eq:gasp})]. To find values for $T$ and $\Teff$, we look for solutions that satisfy Eqs.~(\ref{eq:Sirkocs}) and (\ref{eq:Sirkoradiationequilibrium}) simultaneously.

\begin{figure*}
    \centering
    \includegraphics[width = \textwidth]{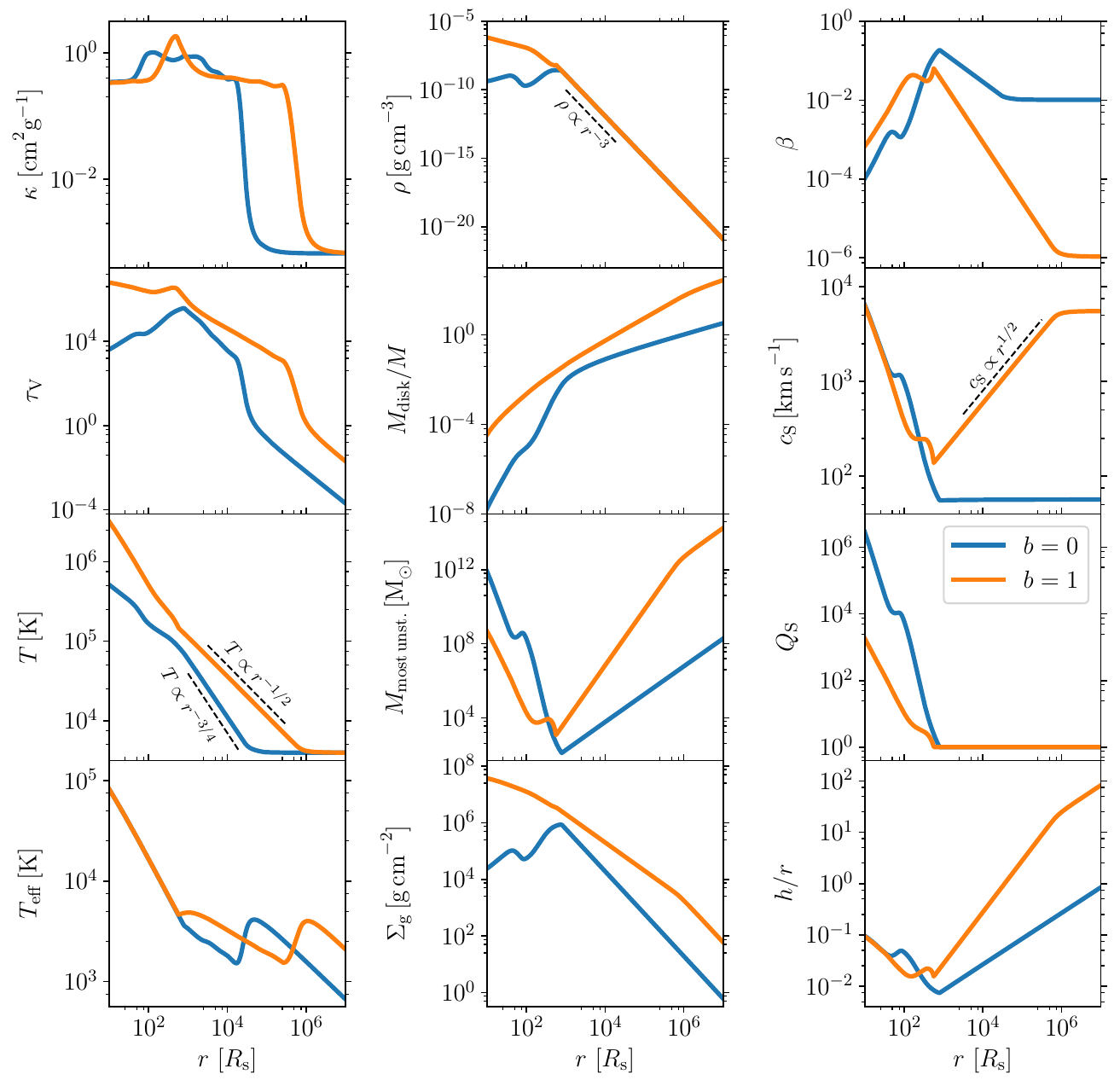}
    \caption{Radial profile in the \SK disc model for the opacity $\kappa$, the mass density $\rho$, the pressure fraction $\beta$, the optical depth $\tauv$, the disc mass $M_{\rm disc}$, the sound speed $c_\mathrm{s}$, the temperature $T$, the most unstable mass $M_{\rm{most} \, \rm{unst.}}$, the Toomre stability parameter $Q_\mathrm{S}$, the effective temperature $\Teff$, the surface mass density $\Sigma_\mathrm{g}$ and the half-thickness $h$. All input values are the same as in Fig.~2 by \protect\cite{2003MNRAS.341..501S}: $M = 10^8 M_{\odot}$, $\alpha = 0.01$, $l_\mathrm{E} = 0.5, \epsilon_{\rm S} = 0.1$. Blue (orange) curves indicate the case where $b=0$ ($b=1$) and the viscosity is proportional to total (gas) pressure. } 
    
    \label{fig:SirkoFig2}
\end{figure*}

\subsubsection{Disc profiles} \label{sec:Sirkoscalings}
Figure~\ref{fig:SirkoFig2} shows the radial profile of some key disc parameters in the \SK model, tailored to reproducing fig.~2 from \SKO. In particular, the figure shows a $10^8 M_{\odot}$ BH 
surrounded by a disc 
with $\alpha = 0.01$, $l_E = 0.5$ and $\epsilon_{\rm S} = 0.1$, presenting both the $b=0$ and $b=1$ cases.

In fig.~2 in \SKO, there are three different solutions for the disc parameters from $r \gtrsim 5 \times 10^5 R_{\rm s}$. Our implementation recovers the same behavior, but we only accept the continuous, high-temperature, low-opacity solution.
For this case, the midplane temperature of the disc remains above $10^3$~K and the opacity drops to $10^{-3}$~cm$^2$g$^{-1}$ in the outer regime, both of which affect the gas and radiation pressure profiles, as reflected in the parameter $\beta$. The transition between the inner and outer disc regimes takes place at $r \approx 10^3 R_\mathrm{s}$, which is consistent with the original results by \SKO.

Figure~\ref{fig:SirkoFig2} also compares the $b=0$ case with the $b=1$ case for the same AGN disc. The difference between the two is that the viscosity is assumed to be proportional to the total and gas pressure, respectively. The $b = 1$ case remains in the optically thick regime out to larger separations, thus also maintaining higher temperatures in those outer regions. For this set of input parameters, the aspect ratio of the disc becomes $>1$ at separations as small as $r \approx 5 \times 10^4 R_\mathrm{s}$, which breaks the thin-disc assumption.

In the outer regime, the density scales as $\rho \propto \Omega_\mathrm{S}^2 \propto r^{-3}$
[see Eq.~(\ref{eq:SirkoQrho})].
Furthermore, the condition $r\gg r_{\rm min}$ implies that $\Mdot '$ is approximately constant. Depending on the value of $b$, one can use Eq.~(\ref{eq:Sirkoviscosityaccretion}) to relate $\rho$, $T$ and $\Omega_\mathrm{S}$. By approximating the system as optically thick ($\tauv \gg 1$), Eq.~(\ref{eq:Sirkoradiationequilibrium}) gives $T^4 \propto \tauv \Teff^4$. Using Eq.~(\ref{eq:Sirkocs}), one can then find simple power-law scalings for most parameters in the outer regime, as long as the opacity $\kappa$ is kept constant. These are shown Fig.~\ref{fig:SirkoFig2} for the $10^3 R_{\rm s} \lesssim r  \lesssim 10^5 R_{\rm s}$ region. In particular, one has $T \propto r^{-3/4}$ for $b = 0$ and $T \propto r^{-1/2}$ for $b = 1$. %
 From Eq. (\ref{eq:Sirkocs}), we find that in the optically thick regime,  $c_\mathrm{s}$ is approximately constant when $b=0$ and proportional to $r^{1/2}$ when $b=1$. %
 At $r \gtrsim 10^5 R_{\rm s}$,  one has $\kappa \ll 1$ and both discs fall back to the optically thin regime. In this case, Eq.~(\ref{eq:Sirkoradiationequilibrium}) gives $T^4 \propto \Teff^4 / \tauv$, which from Eq.~(\ref{eq:Sirkocs}) gives  $T^4 \propto \rho c_\mathrm{s}^2 / \tauv^2$ for both $b=0$ and $b=1$. If we assume $\kappa$ to be constant, then $T$ will also remain constant for all values of $r$.

Figure~\ref{fig:SirkoFig2} shows the ``most unstable mass'' $M_{\rm most \, unst.} \equiv c_{\rm s}^4/G^2\Sigma_{\rm g}$ at a given radius. This is the mass enclosed in protostellar clumps with a characteristic radius $r_{\rm c} = c_{\rm s}^2/G\Sigma_{\rm g}$ \citep{1964ApJ...139.1217T} and corresponds to the maximum mass that can be present in local perturbations and is thus  available for %
 star formation. Figure~\ref{fig:SirkoFig2} shows that, for both the $b=0$ and $b=1$ cases, $M_{\rm most \, unst.}$ has a minimum at $r \approx 10^3 R_{\rm s}$, corresponding to high $\Sigma_{\rm g}$ values and low $c_{\rm s}$ values. Below this radius, where $Q>1$ and star formation ceases, the value of $M_{\rm most \, unst.}$ no longer provides meaningful information.

\subsection{Thompson {et al.} (2005)}

\subsubsection{Modeling strategy}

\TPO proposes an AGN model for which the outer areas of the disc are vertically supported against gravitational collapse by radiation pressure from star formation by-products, dominated in the optically thick regime by dust grains around massive stars. 
The angular momentum transport in the \TP disc is described by global torques instead of a local viscosity mechanism like in the \SK model, which provides rapid radial advection rates in the outer regions of the disc.

In the \TP model, the angular velocity
\begin{equation} \label{eq:ThomOmega}
    \Omega_{\rm T} = \sqrt{\frac{GM}{r^3} + 2 \frac{\sigma^2}{r^2}}
\end{equation}
 is only approximately Keplerian  and includes the effect of the dispersion velocity $\sigma$. The dispersion and the central mass are related by the $M-\sigma$ relation from observations. %
 \TPO used the %
 expression by \cite{2002ApJ...574..740T}, 
while we opted for an updated fit by \cite{2009ApJ...698..198G}
\begin{equation} \label{eq:Msigma}
    \log{\frac{\sigma}{200 \, {\rm km/s}}} =  \frac{1}{4.24} \left(\log{\frac{M}{ M_{\odot}}} - 8.12 \right) \, ,
\end{equation}
which is taken from their full galaxy sample.
We stress that both of these expressions were obtained for surveys of non-AGN galaxies, meaning that they do not appropriately account for selection biases \citep{2017MNRAS.468.4782B, 2017MNRAS.466.4029S,2023A&A...674A.181M}.

The \TP model accounts for the star-formation 
rate per unit area:
\begin{equation} \label{eq:Thometa}
    \dot{\Sigma}_* = \Sigma_\mathrm{g} \Omega_\mathrm{T} \eta \, ,
\end{equation}
which is parametrized using the fraction $\eta$ of the disc dynamical timescale. By means of $\dot{\Sigma}_*$, the \TP model explicitly tracks changes in the accretion rate $\Mdot$ throughout the disc due to star formation. 
The gas accreted onto the central BH is supplied by material outside of a radius $\Rout$ 
at a constant rate $\Mdotout$. As \TPO point out, the AGN disc for the \TP model does not have a clear outer boundary because the gas is expected to be fed to the central BH by the surrounding interstellar medium. Unlike $r_{\rm max}$ in the \SK model, which is a chosen value after which the gas is expected to fragment into stars, here $\Rout$  represents a transition point beyond which the accretion rate is constant and within which the accretion rate varies due to star formation.
Opposite to the \SK case, in the \TP model one integrates from the outer boundary of the AGN disc $\Rout$ down to the inner edge of the disc, here set to $r_{\rm min} = 3 R_\mathrm{s}$.
 
In the \TP model, the  \cite{1964ApJ...139.1217T} stability criterion is written as
 \begin{equation}\label{eq:ThompQ}
    Q_\mathrm{T} = \frac{\kappa_{\Omega} c_\mathrm{s}}{ \pi G \Sigma_\mathrm{g}} \approx \frac{\Omega_\mathrm{T}^2}{\sqrt{2} \pi G \rho} \, ,
\end{equation}
where $\kappa_{\Omega}^2 = 4 \Omega_\mathrm{T}^2 + \der \Omega_\mathrm{T}^2 / \der \ln{r} $ is the epicyclic frequency. To first order in $1/r$, Eq.~(\ref{eq:ThomOmega}) gives $\der \Omega_{\rm T} / \der r \approx -\Omega_T/r$ such that $\kappa_{\Omega} \approx \sqrt{2} \Omega$.
When $Q_\mathrm{T} \gg 1$, we expect conditions to be unfavorable to star formation so that $\dot{\Sigma}_*$ and $\eta$ are close to zero.  In the outer area of the disc where $Q_\mathrm{T} \approx 1$, stellar feedback plays a key role in stabilizing the disc. 

Much like the \SK model,\footnote{For small values of $r$ one has that $\Omega_{\rm T}$ is approximately Keplerian and $Q_{\rm T} \approx Q_{\rm S}$. Since $Q_{\rm T}$ is expected to be $\gg 1$ near the BH, the factor $\sqrt{2}$ is negligible.} the \TP one also has two regimes according to the value of $Q_\mathrm{T}$, see Fig.~\ref{fig:thompson_flowchart}. 
We initialize our numerical root finder at the outer boundary assuming that the disc is optically thick to its own infrared radiation and that $Q_\mathrm{T} = 1$, thus obtaining initial values for $T$, $\rho$ and $\eta$ (see Appendix~\ref{sec:optthick}).

\subsubsection{Non star-forming regime}

For every value of $r$ under consideration, we first assume that there is no star formation and that $Q_\mathrm{T} > 1$. In this case, the accretion rate is constant and thus the value of $\Mdot$ is the same as that of the preceding separation, i.e. $\Mdot (r) = \Mdot(r + \Delta r)$, where $\Delta r$ is the numerical radial resolution. At $r = \Rout$, one has the boundary condition $\Mdot(\Rout) = \Mdotout$. 
The gas ring at cylindrical radius $r$ is assumed to radiate as a black body with effective temperature:
\begin{equation} \label{eq:ThomnostarTeffMdot}
    \sigma_{\rm SB} \Teff^4 = \frac{3 \Omega_\mathrm{T}^2 }{8\pi} \Mdot' \, ,
\end{equation}
which is the same as Eq.~(\ref{eq:SirkoTeffMdot}). The \TP model assumes that the angular momentum in the disc is transported by global torques, so that the radial velocity of the gas $v_{\rm r}$ is a fraction $m_\mathrm{T}$ of the sound speed $c_\mathrm{s}$. The resulting accretion rate is \begin{equation} \label{eq:ThomMdotv}
\Mdot = 4 \pi r \rho h v_{\rm r} = 4 \pi r \rho h m_\mathrm{T} c_\mathrm{s} =  4 \pi r \Omega_\mathrm{T} m_\mathrm{T} \rho h^2 \, ,
\end{equation}
where we have assumed hydrostatic equilibrium, $h = c_\mathrm{s}/\Omega_\mathrm{T}$ [see Eq.~(\ref{eq:Sirkoh})]. Using Eq.~(\ref{eq:ThomMdotv}), one can compute the disc half thickness $h$ as a function of the accretion rate and density.

We then interpolate the opacity tables of our choosing to find the $\kappa (\rho, T)$, which in turn gives us the optical depth $\tauv = \kappa \rho h$ as a function of $T$ and $\rho$. Notably, \TPO use the opacities by \cite{2003A&A...410..611S} which are provided for temperatures up to $T \simeq 10^4$K and extrapolate them to higher temperatures by keeping  $\kappa (\rho, T)$ constant; in the following we refer to this set as the ``Semenov'' opacities. 
In \codename, we instead use the combined set of opacities with values by \cite{2003A&A...410..611S}  up to $T = 10^4$K and then values by \cite{2005MNRAS.360..458B} for higher temperatures.

We look for solutions in $T$ and $\rho$ so that the gas ring is in radiative equilibrium and the sound speed is consistently defined. The condition for radiative equilibrium adopted by \TPO is
\begin{equation} \label{eq:ThomTeff4}
    T^4 = \left(\frac{3}{4} \tauv + \frac{1}{2\tauv} + 1 \right) \Teff^4 \, ,
\end{equation}
which is the same as Eq.~(\ref{eq:Sirkoradiationequilibrium}) but doubled.
The definition of the sound speed $c_{\rm s} = p_{\rm tot}/\rho$ is almost identical to that given in  Eq.~(\ref{eq:Sirkocs}) for the \SK model. The sound speed definition assumes hydrostatic equilibrium and the pressure definitions $p_{\rm gas} = \rho \kb T / m_{\rm U}$, $p_{\rm rad} = \sigma_{\rm SB} \tauv \Teff^4/c$. The additional factor of $2$ in the \TP model's definition of $p_{\rm rad}$ ensures that in the optically thick regime using Eq.~(\ref{eq:ThomTeff4}) gives $p_{\rm rad} \approx 4 \sigma_{\rm SB} T^4 /3c$. 

Solutions for $\rho$ and $T$ are then found by balancing Eqs.~(\ref{eq:Sirkocs}) and (\ref{eq:ThomTeff4}). One can 
then compute  $Q_\mathrm{T}$ once more using Eq.~(\ref{eq:ThompQ}). If  $Q_\mathrm{T}<1$, the ring at radius $r$ is instead assumed to be in the outer star-forming regime.

\subsubsection{Star-forming  regime}

In the case where there is star formation, the accretion rate is no longer constant. Instead, it is calculated numerically by taking the difference between the initial $\Mdotout$ and the integrated accretion rate from star formation down to the current ring:
\begin{align}
\Mdot (r) &= \Mdotout - \int_{r}^{\Rout} 4 \pi r \rho h  \Omega_\mathrm{T} \eta \mathrm{d} r  
\\ &\approx \Mdotout - \sum_{r_j = r}^{\Rout} 4 \pi r_j \rho_j h_j  \Omega_{\mathrm{T,} j} \eta_j \Delta r_j \, ,
\end{align}
where the subscript $j$ denotes that the given parameter is taken at $r = r_j$. Like in the \SK model, we assume marginal stability, i.e. $Q_\mathrm{T} = 1$. Rearranging Eq.~(\ref{eq:ThompQ}) for the mass density yields
\begin{equation} \label{eq:Thomrhoout}
    \rho = \frac{\Omega_\mathrm{T}^2}{\sqrt{2} \pi G} \, .
\end{equation}
The two parameters we are solving for in this case are the temperature $T$ and the star formation fraction $\eta$ of the disc ring. One calculates $h$ from Eq.~(\ref{eq:ThomMdotv}), interpolates the value of $\kappa (T)$ and finds $\tauv (T)$, finally calculating $\Teff^4(T)$ using Eq.~(\ref{eq:ThomTeff4}). 

We now look for solutions in $\eta$ and $T$ that balance the radiated flux
\begin{equation}\label{eq:radbalance_Q1}
    \sigma_{\rm SB} \Teff^4 = \rho h \Omega_\mathrm{T} \eta \epsilon_\mathrm{T} c^2 + \frac{3}{8\pi} \Mdot' \Omega_\mathrm{T}^2 \,,
\end{equation}
which now directly accounts for radiation from stars unlike Eqs.~(\ref{eq:SirkoTeffMdot}) and (\ref{eq:ThomnostarTeffMdot}),
while assuming hydrostatic equilibrium. One has
\begin{equation}\label{eq:hydroeq_Q1}
    \rho h^2 \Omega_\mathrm{T}^2 = \frac{\rho \kb T}{m_\mathrm{U}} +  2 \rho h \eta \Omega_\mathrm{T} \epsilon_\mathrm{T} c \left(\frac{\tauv}{2} + \xi \right) \, ,
\end{equation}
where $\epsilon_\mathrm{T}$ and $\xi$ are free parameters describing the efficiency of star formation in the disc and the radiative efficiency of supernovae, respectively. In this regime, it is expected that the gas will be optically thin, and therefore radiation pressure from supernovae is included through the $\xi$ parameter. \TPO sets $\epsilon_\mathrm{T} = 10^{-3}$ and $\xi=1$. We seek the values of $\eta$ and $T$ that simultaneously solve Eqs.~(\ref{eq:radbalance_Q1}) and (\ref{eq:hydroeq_Q1}).

\subsubsection{Accretion criterion}
Unlike \SKO, the \TPO model presents an accretion rate $\Mdot$ that changes as a function of the radial separation $r$. 
This naturally means that if the accretion rate at the outer boundary is too low, not enough gas is able to reach the central BH to maintain high temperatures and bright AGN luminosities, which are expected to be in the $10^{-3} -0.5 \, L_{\rm Edd}$ range, see \cite{2004ApJ...613..109H, 2006ApJ...648..128K, 2015ApJ...815..129S, 2018ApJ...859..116K}. This introduces a minimum threshold for $\Mdotout$. 
\TPO argue that accretion rates of ${\sim} 1 - 10 \rm\, M_\odot {\rm yr}^{-1}$ at the inner disc boundary $r_{\rm min}$ are sufficient to produce a bright AGN when the central BH mass is ${\sim} 10^9 \rm\, M_\odot$. This is equivalent to a minimum BH accretion rate of $\dot{M} \sim 0.2~\dot{M}_{\rm Edd} = 0.2~L_{\rm Edd}/(0.1 c^2)$ at $r = r_{\rm min}$. Using Eq.~(47) in \TPO, we find that over a wavelength range of $[10^{-8} \, \mathrm{m}, 10^{-3} \, \mathrm{m}]$, setting $0.2~\dot{M}_{\rm Edd}$ gives a disc bolometric luminosity of $2\times 10^{-4} \, L_{\rm Edd}$. 

There is no general expression that relates the accretion rate $\Mdotout$ and outer radius $\Rout$ to the BH mass $M$ that would ensure a bright AGN disc. Nonetheless, we can attempt to find such a relationship by considering how the accretion rate at the outer boundary $\Mdotout$ scales with the size of the disc $\Rout$ and the central BH mass.
\TPO proposes a critical value $\Mdot_\mathrm{c}$, obtained by equating the star formation timescale $\tau_{*} = 1/\eta \Omega$ with the advection timescale $\tau_{\rm adv} = r/v_{\rm r}$ to determine whether enough material reaches the central BH to form a luminous signal.
 From Eq. (\ref{eq:ThomMdotv}) we find %
\begin{equation}\label{eq:Mc}
    \Mdot_\mathrm{c} = 4 \pi r^2 \rho h \eta \Omega_\mathrm{T} \, .
\end{equation}
Together with Eq.~(\ref{eq:Mc}), this result can be used to introduce  a dependence on $\Rout$ and $M$ to $\Mdotout$. 
A BH with $M=10^8 \,\rm M_\odot$ surrounded by a \TP disc that has $\Rout = 95$ pc, $\Mdotout = 320  \,\rm M_\odot {\rm yr}^{-1}$, and $\sigma = 188~{\rm km\, s}^{-1}$ satisfies $\Mdotout > \Mdot_\mathrm{c} (r = \Rout)$ and has an accretion rate near the central BH of ${\approx} 1.93 \,{\rm M}_\odot {\rm yr}^{-1} = 0.74 {\Mdot}_{\rm Edd}$ (giving a disc luminosity of $9.61 \times 10^{-4} \, L_{\rm Edd}$). We use these values to keep the ratio of $\Mdotout / \Mdot_{\rm c}$ constant.
Using Eqs.~(\ref{eq:Thometa}),  (\ref{eq:ThomOmega}), (\ref{eq:Mc}), and assuming the optically thick regime (see Appendix~\ref{sec:optthick}),
one can show that $\Mdot_\mathrm{c} \propto r \sigma^2$. Therefore, we scale $\Mdotout$ with the outer boundary of the disc and the dispersion velocity, i.e.
\begin{equation} \label{eq:Mdotoutscaling}
    \Mdotout = 320 \, {\rm M}_\odot {\rm yr}^{-1} \, \left(\frac{\Rout}{95 {\rm pc}}\right) \left(\frac{\sigma }{188 {\rm kms}^{-1}} \right)^{2} \, .
\end{equation}
However, for high masses, Eq.~(\ref{eq:Mdotoutscaling}) is not enough to fulfill the the $\Mdotout > \dot{M}_{\rm c}$ criterion. The inability of $\dot{M}_{\rm c}$ to accurately predict whether a bright AGN disc is formed is not surprising, as it compares the timescales for only one value of $r$. In \TPO, it is stated that discs with $\Mdotout < \Mdot_\mathrm{c} (r=\Rout)$ cannot form bright AGNs.

We find that using $\Mdot_\mathrm{c}$ from Eq.~(\ref{eq:Mc}) as a threshold is too stringent and often omits signals that produce bright AGN discs. In the following, we use Eq.~(\ref{eq:Mdotoutscaling}) as an initial guess for $\Mdotout$ 
but then make adjustments if the accretion rate is not large enough to form a luminous AGN. Developing a full prescription is left to future work. As a precaution to avoid setting an $\Mdotout$ that is too high, \TP suggests a maximum limit for $\Mdotout$ equal to $\Mdot_{\rm max} = 8 \pi \rho h \sigma^2 r / \epsilon_{\rm T} c = L_{\rm M}/\epsilon_{\rm T} c^2$, where $L_{\rm M}$ is the limiting Eddington-like luminosity below which a galaxy will not have momentum driven winds that are high enough to significantly reduce the gas in the disc \citep{2005ApJ...618..569M}.

Other authors have used different values for $\Mdotout$ and $\Rout$. For instance, \citet{2017MNRAS.464..946S} scale the AGN disc down to a Milky-Way type galaxy, using $M = 3 \times 10^6  \,\rm M_{\odot}$, $\Mdotout = 15 \,\Mdot_{\rm Edd}$, and $\Rout = 10 \,$pc, where the latter %
was motivated by the radius of the dusty tori from AGN disc observations \citep{2004Natur.429...47J, 2013A&A...558A.149B, 2019A&A...632A..61G,  2022Univ....8..356S}. %

\subsubsection{Disc profiles}

\begin{figure*}
    \centering
    \includegraphics[width = \textwidth]{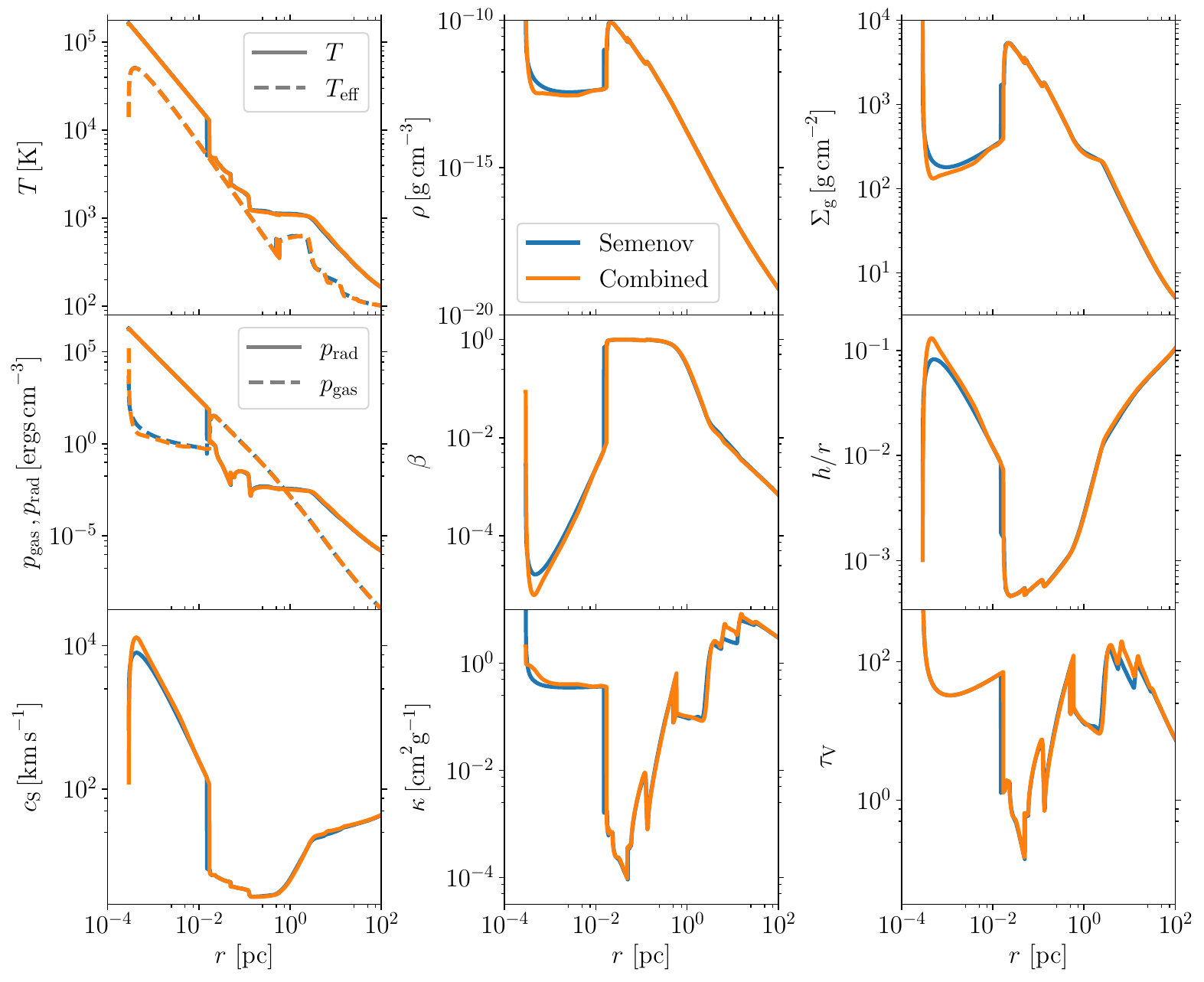}
   \caption{Radial profile in \TP disc model for the temperature $T$, the effective temperature $\Teff$, the mass density $\rho$, the surface mass density $\Sigma_g$, the gas  pressure  $p_{\rm gas}$, the radiation pressure $p_{\rm rad}$, the gas pressure fraction $\beta$, the half-thickness of the disc $h$, the sound speed $c_{\rm s}$, the opacity $\kappa$ and the optical depth $\tauv$. The input values have been chosen to reproduce Fig.~6 in \protect\cite{2005ApJ...630..167T}: $\sigma = 300$~km/s, $\epsilon_\mathrm{T} = 10^{-3}$, $m = 0.2$, $\Mdotout= 320 M_{\odot}$~yr$^{-1}$, and $\Rout = 200$~pc. 
 Models shown in blue use the opacities by \protect\cite{2003A&A...410..611S}, %
 models shown in orange use the combined datasets from \protect\cite{2003A&A...410..611S} and  \protect\cite{2005MNRAS.360..458B}. %
 }
    \label{fig:ThompsonFig6}
\end{figure*}

Figure~\ref{fig:ThompsonFig6} reproduces Fig.~6 in \TPO, assuming either the Semenov opacities (as was done by \citealt{2005ApJ...630..167T}) or the combined opacities. 
The input parameters  are $\sigma = 300$ km~s$^{-1}$, $M \approx 10^9 \, M_{\odot} $  [instead of Eq.~(\ref{eq:Msigma}) we use the $M$-$\sigma$ relation by \cite{2002ApJ...574..740T} as was done by \cite{2005ApJ...630..167T}],
$\Mdotout = 320 M_{\odot}$ yr$^{-1}$, $\Rout = 200$~pc, $m = 0.2$, $\epsilon_\mathrm{T} = 10^{-3}$, and $\xi = 1$. 
Our results shown in Fig.~\ref{fig:ThompsonFig6} are generally in agreement with those by \TPO.

Our implementation results in disc profiles that diverge when the disc is close to the central BH, around $r \sim 10^{-3}$ pc in this case; this follows from the $r \rightarrow r_{\rm min}$ limit in the definition of $\Mdot'$.
We report good agreement between the two opacity implementations, with a noteworthy difference being the presence of the iron opacity bump \citep{2016ApJ...827...10J} at a radius of $r \sim 2 \times 10^{-4}$~pc, seen only for the combined opacities. %
For both sets of opacities, the disc profile presents a sharp  
feature at 
$r \sim 5 \times 10^{-1}$ pc where the temperature becomes high enough to leave the so-called opacity gap
(the dip in $\kappa$ for temperatures $10^3\, {\rm K} \lesssim T \lesssim 10^4\, {\rm K}$, see \citealt{2003A&A...410..611S, 2005ApJ...630..167T}). Figure~\ref{fig:ThompsonFig6} shows that for this set of parameters the disc profile is not sensitive to the choice of opacity tables.

\section{Parameter-space exploration}
\label{sec:disc}

We now present a brief exploration of the phenomenology predicted by the \SKO and \TPO disc models. 

\subsection{Mass dependency}\label{sec:massexplore}

\begin{figure*}
    \centering
    \includegraphics[width = 0.6\textwidth]{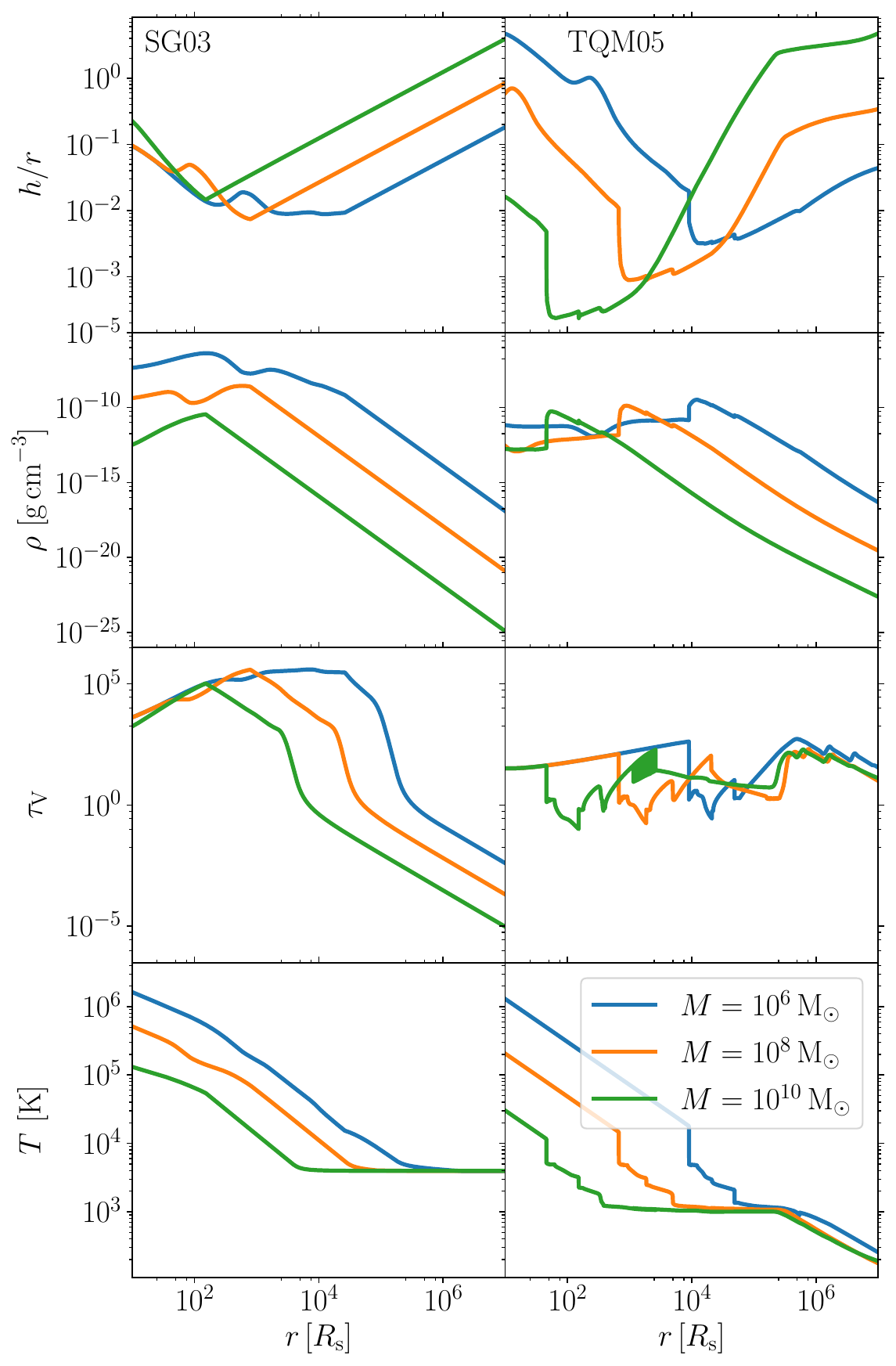}
    \caption{Aspect ratio $h/r$, mass density $\rho$, optical depth $\tauv$ and midplane temperature $T$ as functions of cylindrical radius $r$ for both the \SK (left) and \TP (right) AGN disc models. We vary the central BH mass $M=10^6 \,{\rm M}_\odot$ (blue), $10^8 \,{\rm M}_\odot$ (orange), and  $10^{10} \,{\rm M}_\odot$ (green). For the \SK case, we set $\alpha = 0.01$, $l_\mathrm{E} = 0.5$, and $b=0$. For the \TP case, we set $m = 0.2$, $\epsilon_T = 10^{-3}$ and $\xi = 1$. The outer radius $\Rout$ and outer accretion rate $\Mdotout$ are both scaled with the central BH mass such that $\Rout = 95$ pc and $\Mdotout = 320 \,{\rm M}_\odot {\rm yr}^{-1}$ when $M = 10^8 \,{\rm M}_\odot$, except for the $M = 10^{10} \,{\rm M}_\odot$ disc which has an outer accretion rate set to $\Mdotout = 1.5 \times 10^6 \,{\rm M}_\odot {\rm yr}^{-1}$ .  
    } 
    \label{fig:parammass}
\end{figure*}

We first investigate the behavior of both models as a function of the mass of the central BH. %
Figure~\ref{fig:parammass} compares the \SK and \TP discs profiles of four output parameters, namely the disc height from the midplane $h$, the mass density $\rho$, the optical depth $\tauv$, and the temperature $T$, %
for three central BH masses: $M=10^6$, $10^8$, and $10^{10} \,{\rm M}_{\odot}$. %
These five output quantities can be used to fully reconstruct an AGN disc for both models.
Results are presented using the combined opacity datasets. %

For the \SKO model, we set $\alpha = 0.01$, $l_\mathrm{E} = 0.5$, and only consider the $\alpha$ disc (i.e. $b=0$).  For each disc, we find the solution up to a radius of $10^7 R_\mathrm{s}$, with the $M=10^6 \rm\,M_\odot$ case having a maximum extension of ${\sim} 1$ pc, and the $M = 10^{10} \rm\,M_\odot$ case extending to ${\sim} 1$ kpc. %

The temperature of the \SK disc is higher at small separations for lower masses.
In particular, one has $r\propto R_\mathrm{s} \propto M$ in Fig.~\ref{fig:parammass}, so that $\Omega_\mathrm{S} \propto M^{-1/2}$, and thus $T \propto M^{-3}$ in the inner region, cf. Eq.~(\ref{eq:Sirkoradiationequilibrium}) 
for the optically-thick regime in the \SK model.  
In the outer regions of the \SK disc, all three models have the same temperature $T \approx 7.5 \times 10^3 K$, which is reached at the separation where the disc becomes optically thin ($\tauv < 1$). 
At large radii, if the disc is dominated by radiation pressure and the gas is optically thin ($\Teff^4 \propto \tauv T^4$), then from Eq.~(\ref{eq:Sirkocs}) we find that $c^2_\mathrm{s} \propto \tauv^2 T^4 / \rho$. If $\kappa$ is independent of $r$, then $\tauv \propto \rho h$, which in hydrostatic equilibrium gives a constant $T$ independent of both $r$ and~$M$.

Figure \ref{fig:parammass} shows that the density $\rho$ is lowest when the central BH mass is highest, with $\rho \propto M^2$ in the inner region of the \SK disc.  The model with $M=10^{10} \rm\,M_{\odot}$ presents the thickest \SK disc, reaching $h/r > 1$ at $r \gtrsim 10^6 R_{\rm s}$; this is outside the regime of validity of our equations but only applies for large radii suggesting a diffuse envelope of gas around the AGN disc. %

The \TPO model shown in Fig.~\ref{fig:parammass} uses $m_{\rm T} = 0.2$, $\epsilon_{\rm T} = 10^{-3}$ and $\xi = 1$. In Fig.~\ref{fig:parammass}, we linearly scale the outer boundary of the disc $\Rout$ using the Schwarzschild radius so that $\Rout = 10^7 R_{\rm s}$ for all three BH masses.
We calculate $\Mdotout$ using Eq.~(\ref{eq:Mdotoutscaling}) for all three BH masses, but find that for the $M=10^{10} \,{\rm M}_\odot$ case the scaled $\Mdotout$ does not satisfy the $\Mdotout > \dot{M}_{\rm c}$ condition and the disc profile looks significantly different from the AGN discs with smaller masses (the height ratio $h/r$ monotonically decreases and the temperature in the disc does not reach $10^4$ K). Instead, we opt for $\Mdotout = 1.5 \times 10^6 \,{\rm M}_\odot {\rm yr}^{-1}$  %
when  $M=10^{10} \,{\rm M}_\odot$ instead. Equation~(\ref{eq:Mdotoutscaling}) gives $\Mdotout = 0.37 \,{\rm M}_\odot {\rm yr}^{-1}$ when $M = 10^6 \,{\rm M}_\odot  $ and $\Mdotout = 322 \,{\rm M}_\odot {\rm yr}^{-1}$ when $M = 10^8 \,{\rm M}_\odot $.
The AGN disc with $M=10^8 \,{\rm M}_{\odot}$ has an outer boundary of $100\,$pc, which is about half the size of the model shown in Fig.~\ref{fig:ThompsonFig6}.

The $M=10^6 \,{\rm M}_{\odot}$ case in Fig.~\ref{fig:parammass} shows an AGN disc with an outer boundary $\Rout \approx 1\,$pc and a BH accretion rate $0.37 \, {\rm M}_\odot {\rm yr}^{-1}  $. Its accretion rate $\Mdot$ is higher than both the star formation rate and $\Mdot_{\rm c}$ for all values of $r$, leading to  temperatures as large as $T \sim 10^6 \rm\, K$ at $r = r_{\rm min}$ and a disc luminosity of $2 \times 10^{-5} \, L_{\rm Edd}$. The radiation pressure in such a high-temperature region leads to a thick disc, with $h/r > 1$ below $r \sim 5 \times 10^2 R_{\rm s}$. At this aspect ratio, the thin-disc approximation no longer applies and caution must be applied when interpreting our results. In order to reduce $h/r$ in the inner regime, one can decrease $\Mdotout$  or decrease $m_{\rm T}$.

We find that the model with $M=10^{10}\,\rm M_{\odot}$  also reaches $h/r > 1$ but at $r > 10^5\, R_{\rm s}$. This is due to a combination of low densities, a large optical depth, and a large accretion rate which all increase the radiation pressure at the outer boundary. The $M=10^{10} \,\rm M_{\odot}$ AGN disc extends out to $10$~kpc and has an accretion rate of ${\sim} 10 \,\rm M_{\odot} {\rm yr}^{-1} = 0.04 ~\Mdot_{\rm Edd}$ at $r = r_{\min}$, giving a disc luminosity of $0.07 \, L_{\rm Edd}$. For the \TP model with $M=10^{10}\,\rm M_{\odot}$, the optical depth   $\tau_{\rm V}$ shows oscillations at $r\sim 200 R_{\rm s}$ (see Fig.~\ref{fig:parammass}) which are due to the model switching between the inner and outer regimes back and forth when close to the $Q_{\rm T} = 1$ boundary.

\subsection{Input parameters}\label{sec:inputparamexplore}

\begin{figure*}
    \centering
    \includegraphics[width = 0.6\textwidth]{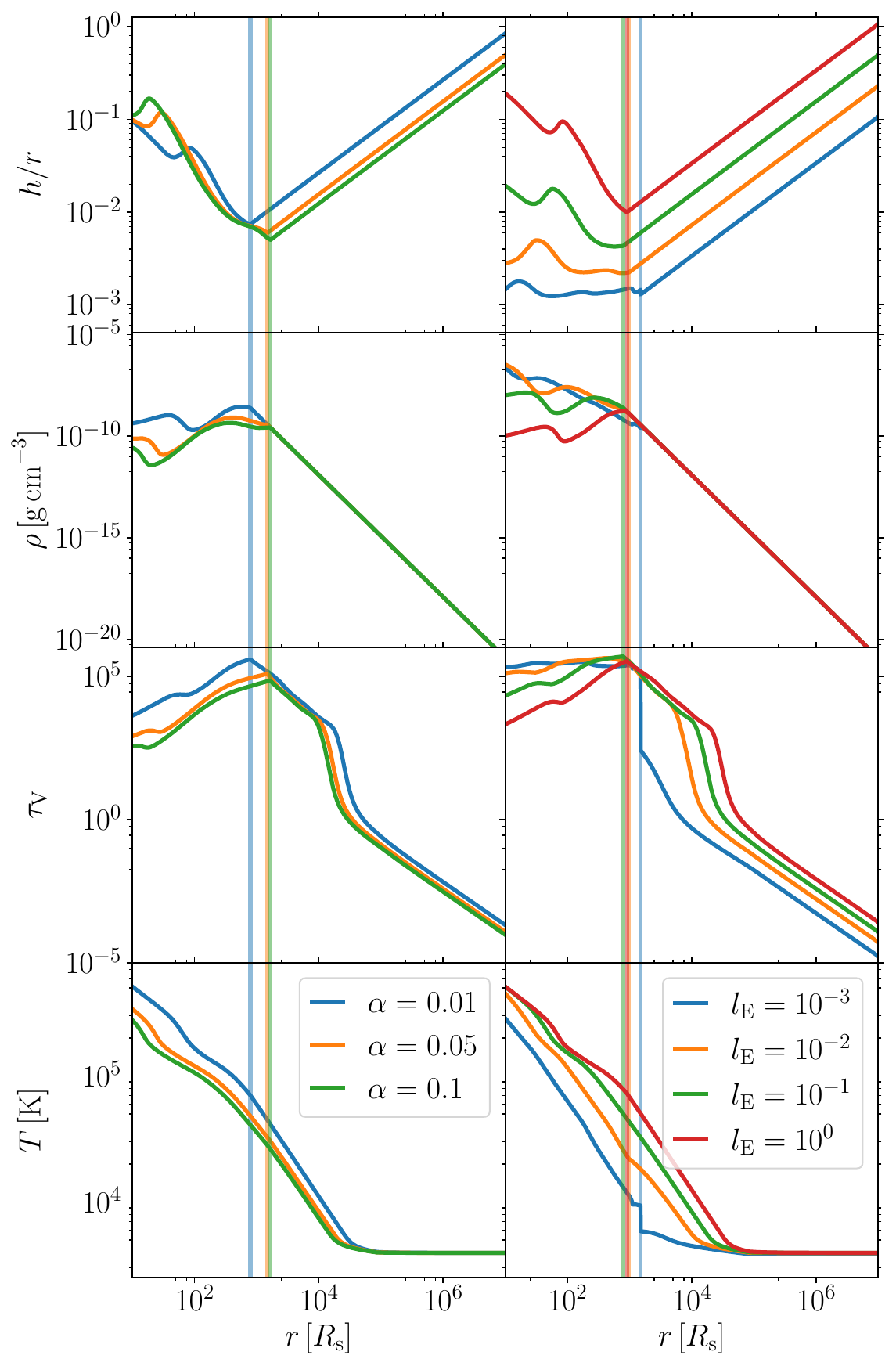}
    \caption{Model variations for the \SK model, showing in particular the aspect  ratio $h$, the midplane mass density $\rho$, the optical depth $\tauv$ and the midplane temperature $T$. For both columns, we set $M = 10^8 \,{\rm M}_\odot$ and $b=0$. In the left column, we consider AGN discs with an Eddington fraction $l_\mathrm{E} = 0.5$ and vary the viscosity with $\alpha = 0.01$ (blue), $\alpha = 0.05$ (orange) and $\alpha = 0.1$. In the right column, we consider AGN discs where $\alpha = 0.01$, and vary the Eddington ratio  $l_{\rm E} = 0.001$ (blue), $l_{\rm E} = 0.01$ (orange), $l_{\rm E} = 0.1$ (green) and $l_{\rm E} =1$ (red). For each disc instance, the radius at which $Q_{\rm S} = 1$ is marked by a vertical line.
 }
    \label{fig:Sirkoparams}
\end{figure*}

The \SK model has five input parameters: the mass of the central BH $M$, the luminosity ratio $l_{\rm E}$ (or alternatively the accretion rate $\Mdot$), the disc viscosity $\alpha$, the BH radiative efficiency $\epsilon_\mathrm{S}$, and the pressure flag  $b=0,1$. We consider a fiducial model with $M=10^8 \rm\,M_\mathrm{\odot}$\, $\epsilon_\mathrm{S} = 0.1$, $\alpha = 0.01$, $l_\mathrm{E} = 0.5$ and $b = 0$. Of these parameters, Fig.~\ref{fig:Sirkoparams} explores the effect of varying $\alpha$ and $l_{\rm E}$.

The density $\rho$ in the outer regime is largely independent of $\alpha$ and $l_\mathrm{E}$. The \cite{1973A&A....24..337S} parameter $\alpha$ relates the viscosity to pressure and accretion, cf. Eq.~(\ref{eq:Sirkoviscosityaccretion}). A larger $\alpha$ in the \SKO model implies a lower density and lower temperature in the inner regime, cf.  Fig.~\ref{fig:Sirkoparams}. In the outer regions, the density is independent of the viscosity and thus independent of $\alpha$.

We vary the Eddington ratio from $l_{\rm E} = 10^{-3}$ to $l_{\rm E} = 1$, capturing the range of observed AGNs \citep{2004ApJ...613..109H, 2006ApJ...648..128K, 2015ApJ...815..129S, 2018ApJ...859..116K}.
The Eddington ratio parameterizes the accretion rate, which plays a key role in the disc dynamics at all radial distances from the BH. Scaling relations in the optically thick regime far from the disc (see Sec.~\ref{sec:Sirkoscalings}) indicate that the \SK model maintains a constant temperature and density at $r \gtrsim 10^5 R_\mathrm{s}$. Higher %
 accretion rates leads to higher effective temperatures [Eq.~(\ref{eq:SirkoTeffMdot})], %
  higher disc temperatures overall [Eq.~(\ref{eq:Sirkoradiationequilibrium})],  %
and higher total pressure in the disc [Eq.~(\ref{eq:Sirkoviscosityaccretion})], %
 which also implies that $h$ must be higher to maintain hydrostatic equilibrium. %

\begin{figure*}
    \centering
    \includegraphics[width = 0.6\textwidth]{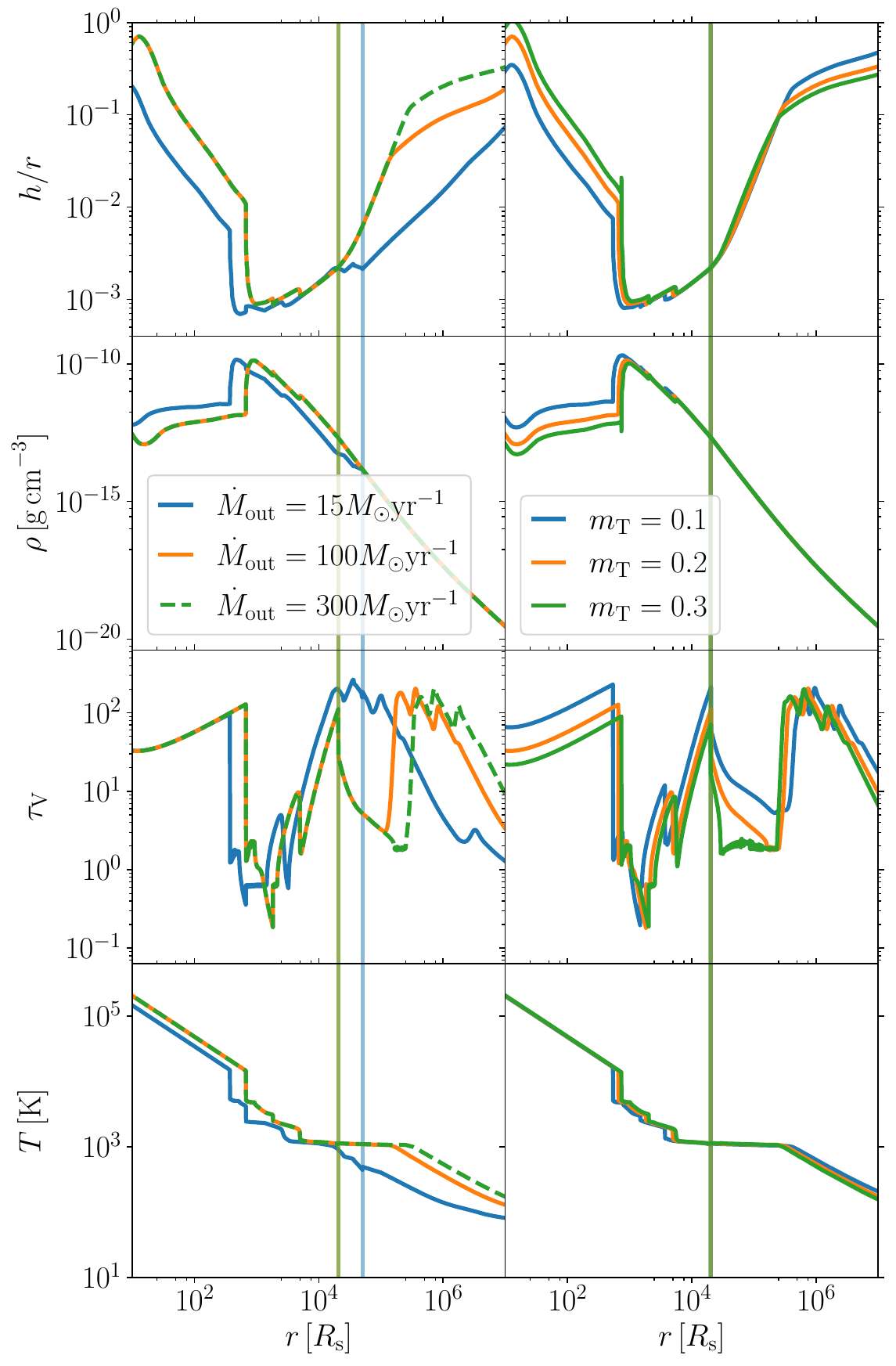}
    \caption{
    Model variations for the \TP model, showing in particular the aspect ratio $h$, the midplane mass density $\rho$, the optical depth $\tauv$ and the midplane temperature. For both columns, we set $M = 10^8 \,{\rm M}_\odot$, $\epsilon_{\rm T} = 10^{-3}$, $\xi = 1$ and $\Rout = 200$ pc.
    In the left column, we consider AGN discs with a global torque efficiency of $m_{\mathrm{T}}= 0.2$ and vary the accretion rate $\Mdotout = 15 \,{\rm M}_\odot {\rm yr}^{-1}$ (blue), $\Mdotout = 100 \,{\rm M}_\odot {\rm yr}^{-1}$ (orange)
   and $\Mdotout = 300 \,{\rm M}_\odot {\rm yr}^{-1}$ (green, dashed). The $\Mdotout = 300 \,{\rm M}_\odot {\rm yr}^{-1}$ case is dashed to show that parameter profiles are identical to the those of the $\Mdotout = 100 \,{\rm M}_\odot {\rm yr}^{-1}$ case close to the central BH. In the right column, we consider AGN discs with an outer accretion rate
    $\Mdotout \simeq 312 \,{\rm M}_\odot {\rm yr}^{-1}$  
    and vary the global torque efficiency 
  $m_{\rm{T}} = 0.1$ (blue),
    $m_{\rm{T}}=0.2$ (orange), and $m = 0.3$ (green).  For each disc instance, the radius at which $Q_{\rm T}= 1$ is marked by a vertical line.
}
    \label{fig:Thompsonparams}

\end{figure*}

The \TP model has six input parameters: the mass $M$ of the SMBH from which we get the velocity dispersion $\sigma$ using Eq.~(\ref{eq:Msigma}),
the star formation efficiency $\epsilon_\mathrm{T}$, the efficiency of angular momentum transport $m_\mathrm{T}$ in the disc, the supernovae radiative fraction $\xi$, the outer boundary of the disc $\Rout$, and the accretion rate at this outer boundary $\Mdotout$. Figure~\ref{fig:Thompsonparams} assumes a fiducial model  with $M=10^8 \,{\rm M}_\odot$, $\Rout = 10^7 R_\mathrm{s}$, $\epsilon_\mathrm{T} = 10^{-3}$, $\xi = 1$,  $m_{\rm T} = 0.2$ and $\Mdotout
 \approx 312  \,{\rm M}_\odot {\rm yr}^{-1}$ from Eq.~(\ref{eq:Mdotoutscaling}). Starting from this set of parameters, we explore how the disc profile changes when varying either $\Mdotout$ or~ $m_\mathrm{T}$.

 We consider three values of the accretion rate: $\Mdotout =15$, $100$, $300 \,{\rm M}_\odot {\rm yr}^{-1}$. The lowest accretion rate considered, $\Mdotout = 15 \,{\rm M}_\odot {\rm yr}^{-1}$, falls below the critical accretion rate $\dot{M}_\mathrm{c} \approx 21 \,{\rm M}_\odot {\rm yr}^{-1}$ from Eq.~(\ref{eq:Mc}) at $r = \Rout$. According to this criterion, this model should not produce an AGN that is sufficiently bright. At $r = r_{\rm min}$, the accretion rate for the $\Mdotout = 15 \,{\rm M}_\odot {\rm yr}^{-1}$ case is $\sim 0.58 \,{\rm M}_\odot {\rm yr}^{-1} = 0.22 \,\Mdot_{\rm Edd}$, which is below the $1-10 \,{\rm M}_\odot {\rm yr}^{-1}$ threshold indicated by \TPO. For this case, the disc luminosity is $1.8 \times 10^{-4} \, L_{\rm Edd}$, which still falls in the range of Eddington ratios one might expect for AGN discs. This further shows that $\Mdot_{\rm c}$ is too strict a criterion for determining whether a \TP disc forms an AGN.
The disc with such a low accretion rate has a different structure compared to the other two cases, with temperatures that are typically lower. As illustrated in Fig.~\ref{fig:Thompsonparams}, these low temperatures lead to low radiation pressure that fails to effectively counteract the vertical collapse of the disc and thus lower $h/r$ values. On the other hand, for cases where the outer accretion rates clear the $\dot{M}_\mathrm{c}$ criterion, we find that the profiles become identical when in the inner, non-star forming regime, see the region left of the $Q_{\rm T}=1$ line in Fig.~(\ref{fig:Thompsonparams}). %
For these cases, the advection timescale and star formation timescale reach an equilibrium at the opacity gap ($\tau_{\rm adv} = \tau_{*}$ when $T \approx 10^3 \, {\rm K}$). This leads to discs of the same temperature, density, aspect ratio and accretion rate ($\Mdot = 2.23 \,{\rm M}_\odot {\rm yr}^{-1}$ at $r = r_{\rm min}$ for both discs).

The global torque efficiency parameter $m_\mathrm{T}$ is strongly correlated to the behaviour of the disc for all radial distances. Much like $\alpha$ for the \SK model,
here $m_\mathrm{T}$ parametrizes the relationship between the angular momentum transport and the accretion rate, cf. Eq.~(\ref{eq:ThomMdotv}). In the outermost regions of the disc, where the accretion rate $\Mdot \approx \Mdotout$ is roughly constant and $h/r \sim 1$, the total pressure is inversely proportional to $m_{\rm T}$, see Eq.~(\ref{eq:ThomMdotv}) and the definition of $c_\mathrm{s}$. This gives a thinner \TP disc, with lower values of $h/r$ for higher values of $m_\mathrm{T}$ at the outer boundary. The density is constant in the marginally stable outer region because of Eq.~(\ref{eq:ThompQ}), but the low total pressure causes some temperature deviations at $r \approx 10^7 R_\mathrm{s}$ for each disc we consider. These variations contribute to different initial conditions in $\tauv$ 
for each value of $m_\mathrm{T}$. The \TP  disc has similar behavior for all three $m_\mathrm{T}$ values once the solutions enter the opacity gap at $r \approx 10^5 R_\mathrm{s}$, though differences in the optical depth impact the disc profiles at small values of $r$. In the innermost regions of the disc, we find that high $m_\mathrm{T}$ values lead to thick, low density discs due to low radiation pressure (which is proportional to $\tauv$ by definition).

\section{Disc migration}
\label{sec:migration}

Migration in gas discs was first proposed based on formulae that considered how spiral-wave structures can be sustained in galaxies \citep{1973StAM...52....1F, 1979ApJ...233..857G, 1979MNRAS.186..799L}. An object orbiting in a gas disc
 exchanges angular momentum with its surroundings, leading to changes in its orbit and thus a net radial migration in the disc \citep{2020apfs.book.....A}. 
These migration torques were predicted by \citet{1979ApJ...233..857G}, described for planets in protoplanetary discs by \citet{1997Icar..126..261W}, improved upon by \cite{2006A&A...459L..17P, 2010MNRAS.401.1950P}, studied for the case of planets by \cite{2000MNRAS.318...18N, 2002ApJ...565.1257T, 2006A&A...459L..17P, 2008A&A...487L...9K, 2010MNRAS.401.1950P, 2010ApJ...715L..68L}, and  extended to the AGN case by \cite{1991MNRAS.250..505S, 1993ApJ...409..592A, 2003astro.ph..7084L, 2011MNRAS.417L.103M, 2012MNRAS.425..460M, 2016ApJ...819L..17B, 2023MNRAS.521.4522D}. 
A key phenomenon emerging from these studies is the potential occurrence of migration traps --- locations in the gas disc where the net radial migration torque is zero. Depending on the mass ratio between the migrator, the central object or the disc, the migrator may clear a gap (referred to as Type~II migration) or deplete material at the migration trap without clearing a gap (Type~I migration) \citep{1997Icar..126..261W}. In this work, we only consider the 
case  of Type~I migration.
Migration traps are an effective mechanism for merging stellar-mass BH binaries embedded in AGN discs, especially in a hierarchical manner (\citealp{2012MNRAS.425..460M, 2016ApJ...819L..17B, 2019PhRvL.123r1101Y,2023PhRvD.108h3033S,2023arXiv231118548V}, see \citealt{2021NatAs...5..749G} for a review). Earlier works by \cite{2016ApJ...819L..17B} and \cite{2024MNRAS.530.2114G} showed that the the location of migration traps does not depend on the properties of the migrating object. The location of these migration traps turns out to be a non-trivial function of the AGN disc parameters, ultimately set by the complex interplay of the gradients of the surface density $\Sigma_\mathrm{g}$ and temperature $T$.
Migration traps are thus an ideal context to showcase our implementation of the \SKO and \TPO disc models.  Table \ref{tab:paramsmigration} summarizes all parameters used for this section.  

\subsection{Torque implementation}

In particular, we apply our AGN disc models to the methods by \citet{2024MNRAS.530.2114G}, adopting their migration torque and thermal torque expressions. \citet{2024MNRAS.530.2114G} use a simpler AGN disc model where profiles are power laws in $M$, $r$ and accretion rate $\Mdot$. 
Their discs are relatively similar to the \SK models with $M = 10^6 M_\odot$ and $\alpha = 0.01$.
When using migration torques by \citet{2010MNRAS.401.1950P} which assume the disc is locally isothermal, \citet{2024MNRAS.530.2114G} report the existence of migration traps. However, migration traps disappear when considering the updated migration torque formulas by \citet{2017MNRAS.471.4917J}. \citet{2024MNRAS.530.2114G} then add a new type of migratory torque, namely the thermal torque by \cite{2017MNRAS.472.4204M}, and find that migration traps are able to form in their AGN disc model once more. We apply the same methodology and formulas to our more complex AGN models.

\begin{table}
\centering
  \caption{Summary of the parameter entering our treatment of disc migration explored in Sec.~\ref{sec:migration}.
  \label{tab:paramsmigration} }
\begin{tabular}[c]{| l | c |}

 \hline
 \multicolumn{1}{|c|}{\textbf{Symbol}} & \multicolumn{1}{c|}{\textbf{Definition}} \\ \hline
  $m_{\rm BH}$ & Mass of the migrating object\\
   $q$ & Mass ratio between migrator and central BH\\
 $\Gamma_0$ & Normalization migration torque\\
  $\Gamma_{\rm I}$ & Type I migration torque\\
  $\gamma$ & Adiabatic index \\
    $C_{\rm L}$ & Lindblad torque\\
      $\chi$ & Thermal diffusivity of the disc \\
  $\Gamma_{\rm therm}$ & Thermal torque\\
    $x_{\rm c}$ & Corotation radius of the migrator\\
     $\lambda$ & Size of the thermal lobes\\
      $L$ & Migrator luminosity from thermal heating\\
       $L_{\rm c}$ & Critical migrator luminosity\\
      
 \hline
 \end{tabular}

 \end{table}

Migration induces two over-dense spiral arms in the disc. Each arm will produce a torque acting on the migrating object with a magnitude \citep{1993Icar..102..150K}
\begin{equation}
    \Gamma_0 = q^2 \Sigma_\mathrm{g} r^4 \Omega^2 \left(\frac{h}{r}\right)^{-3} \,,
\end{equation}
where  $q \equiv m_{\rm BH}/M$ is the BH mass ratio and $\Omega$ is equal to either $\Omega_{\rm S}$ or $\Omega_{\rm T}$ depending on the AGN disc model. The net torque $\Gamma_{\rm I}$ acting on the migrator in a locally isothermal limit is given by  \citep{2010MNRAS.401.1950P}:
\begin{equation} \label{eq:MigrationPaardekooper}
    \Gamma_{\rm I} = \left( -0.85 + 0.9 \frac{\der \ln \Sigma_\mathrm{g}}{\der \ln r} + \frac{\der \ln T}{\der \ln r} \right) \frac{h}{r} \, \Gamma_0 \, .
\end{equation}
\citet{2017MNRAS.471.4917J} updates the migration torque formula %
to %
\begin{equation} \label{eq:MigrationJimeneztot}
    \Gamma_{\rm I} = \left[ C_\mathrm{L} + \left( 0.46 + 0.96 \frac{\der \ln \Sigma_\mathrm{g}}{\der \ln r} - 1.8 \frac{\der \ln T}{\der \ln r}  \right)\gamma^{-1} \right] \frac{h}{r}\, \Gamma_0 \, ,
\end{equation}
where $\gamma=5/3$ is the adiabatic index. The parameter 
\begin{equation}
    C_\mathrm{L} = \left(-2.34 -0.1 \frac{\der \ln \Sigma_\mathrm{g}}{\der \ln r} + 1.5 \frac{\der \ln T}{\der \ln r} \right) f_\gamma \left(\frac{\chi}{h^2 \Omega}\right) \, ,
\end{equation}
is the Lindblad torque where %
\begin{equation}
    f_\gamma (x) = \frac{(x/2)^{1/2} + 1/\gamma}{(x/2)^{1/2} + 1} \, 
\end{equation}
is a function that adds a dependence on the thermal diffusivity for the Lindblad torque and can be approximated to $1/\gamma$ in the case where the diffusivity is small
\citep{2010ApJ...723.1393M}. The thermal diffusivity of the disc  is defined as
\begin{equation}
    \chi = \frac{16 \gamma (\gamma -1 )\sigma_\mathrm{SB} T^4}{3 \kappa \rho^2 h^2 \Omega^2} \, .   
\end{equation}

The thermal torque $\Gamma_{\rm therm}$  originates from the temperature build-up around the migrating object due to lack of heat release during its orbital evolution. If heat is trapped around the migrator, two cold and dense lobes are formed in the disc, which leads to inward migration \citep{2014MNRAS.440..683L}. If the migrator is instead able to release heat back into the disc around it, two hot and under-dense lobes form, leading to outward migration \citep{2015Natur.520...63B}. The total heating torque
is
 \citep{2017MNRAS.472.4204M}
\begin{equation} \label{eq:Migrationtherm}
    \Gamma_\mathrm{therm} = 1.61 \frac{\gamma - 1}{\gamma} \frac{x_{\rm c}}{\lambda} \left( \frac{L}{L_\mathrm{c}} - 1 \right) \Gamma_0
\end{equation}
where $x_\mathrm{c}$ is the corotation radius of the migrating object, $\lambda$ is the typical size of the lobes, and $L$ is the luminosity generated by the migrator through thermal heating, and  %
\begin{equation}
    L_\mathrm{c} = \frac{4 \pi G q M \rho}{\gamma} \chi
 \end{equation}
 is the critical luminosity. If $L = L_\mathrm{c}$, the hot and cold torques acting on the migrator balance out and $ \Gamma_\mathrm{therm}=0$. 
 We approximate the luminosity of the migrator, $L$, to be its Eddington luminosity [see Eq.~(\ref{eq:Ledd}), replacing the mass $M$ with the mass of the migrator $m_{\rm BH}$]. 
 The size of the lobes
 $\lambda$ is given by %
 \citep{2024MNRAS.530.2114G}
 \begin{equation}
     \lambda = \sqrt{\frac{2 \chi}{3 \gamma \Omega}} \, ,
 \end{equation}
 and the corotation radius is %
 \citep{2024MNRAS.530.2114G}:
 \begin{equation}
     x_{\rm c} = - \frac{h^2}{3 \gamma r} \frac{\der \ln{p_{\rm tot}}}{\der \ln{r}}\, .
 \end{equation}
 We approximate $ \der \ln p_{\rm tot} / \der \ln r$ by combining the equation for vertical hydrodynamical equilibrium $p_{\rm tot} \approx \rho h^2 \Omega^2$  
 with the definition of the sound speed $\cs^2 = h^2 \Omega^2$, %
resulting in $\der p_{\rm tot} / \der r \approx \rho \cs^2 / r$. 

The thermal torque given by Eq.~(\ref{eq:Migrationtherm}) is expected to diminish in optically thin discs. Following \citet{2024MNRAS.530.2114G}, we multiply Eq.~(\ref{eq:Migrationtherm}) by a factor of $1 - \exp{-\lambda \tauv/h }$. Additionally, when the mass of the migrator exceeds the thermal mass $m_{\rm th}$, the thermal torque will be reduced \citep{2021MNRAS.507.3638G}. The thermal mass is defined by:
\begin{equation}
    \frac{m_{\rm th}}{m_{\rm BH}} = \frac{\chi}{c_{\rm s} R_{\rm B}} \, ,
\end{equation}
where $R_{\rm B}$ is half the Bondi radius
\begin{equation}
    R_{\rm B} = \frac{G m_{\rm BH}}{c_{\rm s}^2} \, .
\end{equation}
In the regions where $h < R_{\rm B}$ we use the disk height $h$ in place of half the Bondi radius $R_{\rm B}$. To correct for the critical thermal mass, we split Eq.~\ref{eq:Migrationtherm} into its heating component (the positive $L/L_{\rm c}$ term) and its cooling component (the negative term), which we label as $\Gamma_{\rm therm, \, hot }$ and $\Gamma_{\rm therm, \, cold }$ respectively. The total thermal torque is described by Eq.~\ref{eq:Migrationtherm} unless $\mu_{\rm th} \equiv m_{\rm th} / m_{\rm BH} < 1$. In regions of the disc where $\mu_{\rm th} < 1$, the thermal torque is instead given by:
\begin{equation}
    \Gamma_{\rm therm} = \Gamma_{\rm therm, \, hot} \frac{4 \mu_{\rm th}}{1 + 4 \mu_{\rm th}} + \Gamma_{\rm therm, \, cold} \frac{2 \mu_{\rm th}}{1 + 2 \mu_{\rm th}}
\end{equation}
which is an approximation of numerical fits detailed in \citet{2020MNRAS.495.2063V} and used in \citet{2021MNRAS.507.3638G, 2024MNRAS.530.2114G}.

\begin{figure*}
    \centering
    \includegraphics[width = 0.85\textwidth]{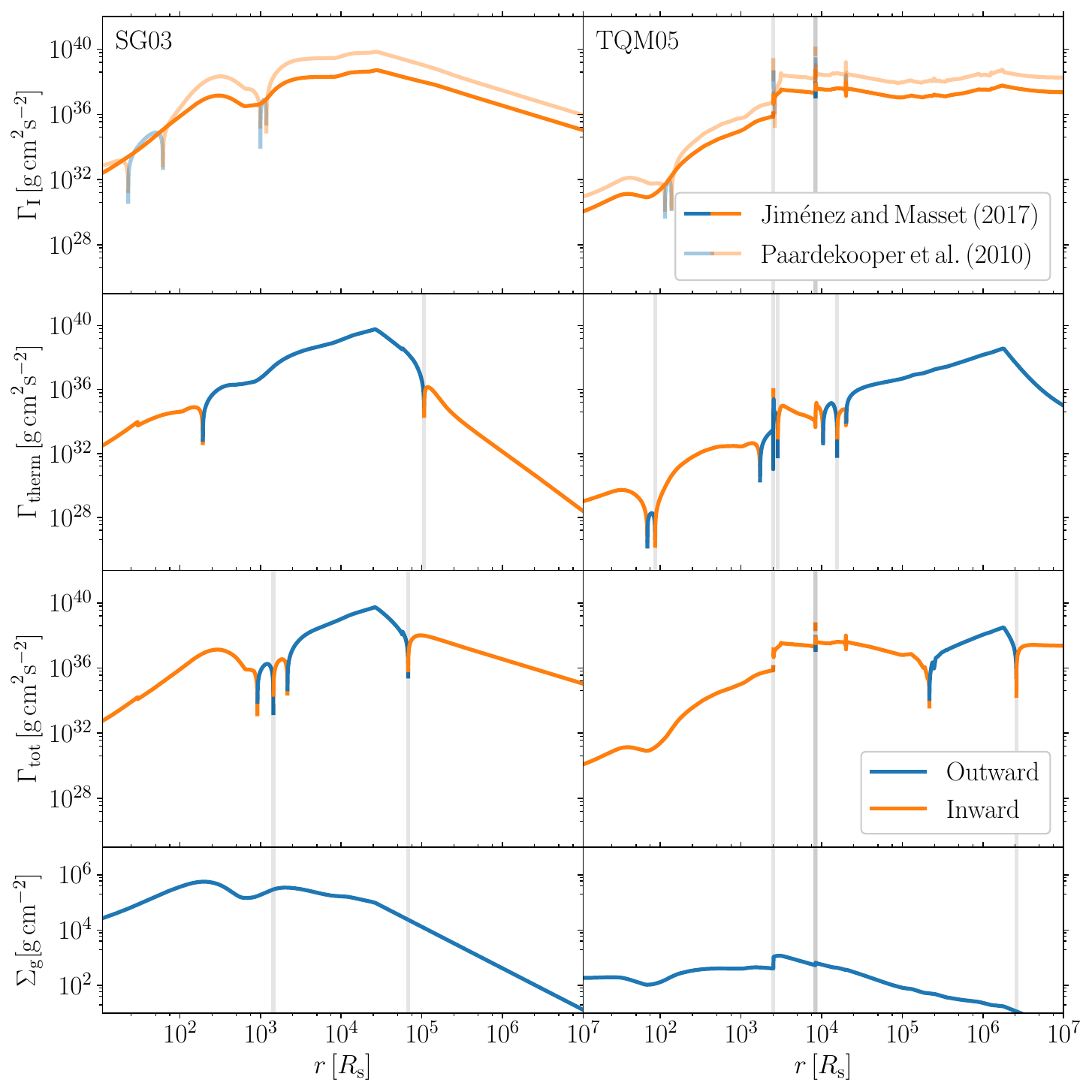}
    \caption{The absolute values of the migration torques for a $m_{\rm BH} = 10 \,{\rm M}_\odot$ BH orbiting a $M=10^6 M_{\odot}$ central BH in  AGN discs. The left panels show torque profiles for a \SK disc with $\epsilon_\mathrm{S} = 0.1$, $\alpha = 0.01$, $l_\mathrm{E} = 0.5$ and $b = 0$. The right panels show torque profiles for a \TP disc with $\epsilon_\mathrm{T} = 10^{-3}$, $\xi = 1$, $m = 0.2$, $\Rout = 10^7 R_\mathrm{s}$ and $\Mdotout = 1.5\times 10^{-2} \, M_{\odot} \, {\rm yr}^{-1}$. %
    Migration torques, thermal torques, and their combination are shown in first three rows from the top, respectively. The bottom panel shows the midplane surface density of the disc for each case. For the Type I migration torques considered in the top row, we show both results using prescriptions  by both \citet{2010MNRAS.401.1950P} (light curves) and \citet{2017MNRAS.471.4917J} (heavy curves). Colors indicate the sign of the torque, with blue referring to inward migration (i.e. positive torques) and orange referring to outward migration (i.e. negative torques). Vertical grey lines indicate the migration traps for all torque prescriptions except for \citet{2010MNRAS.401.1950P} in the top panel. 
  }
  
    \label{fig:migrationtorques}
\end{figure*}

\subsection{Migration traps}

Figure \ref{fig:migrationtorques} shows the  migration-torque profiles for a $M=10^6M_{\odot}$ SMBH in both the \SKO and \TPO model.
We use $\epsilon_\mathrm{S} = 0.1$, $\alpha = 0.01$, $l_\mathrm{E} = 0.5$ and $b = 0$ for the \SK AGN disc and $\Rout = 10^7 R_\mathrm{s}$, $\epsilon_\mathrm{T} = 10^{-3}$, $\xi = 1$,  $m_{\rm T} = 0.2$, and $\Mdotout = 1.5 \times 10^{-2} \, M_{\odot} \, {\rm yr}^{-1}$ %
for the \TP model. The outer accretion rate $\Mdotout$ was set smaller than the value given by Eq.~(\ref{eq:Mdotoutscaling}) in order to enforce $h/r <1$ throughtout the disc, unlike the small mass case in Fig.~\ref{fig:parammass}.
Identifying a migration trap corresponds to regions of the disc where the net migration torque is zero and goes from negative (i.e. inward migration) to positive (i.e. outward migration) as $r$ increases.
 
The top panel shows the migration torque using both Eq.~(\ref{eq:MigrationPaardekooper}) by \citet{2010MNRAS.401.1950P} and  Eq.~(\ref{eq:MigrationJimeneztot}) by \citet{2017MNRAS.471.4917J}.
When using the former, 
we find migration traps at $r\approx 22 R_{\rm s}$ and 
$r\approx 10^3 R_{\rm s}$ for the \SK model, %
which is in line with the results reported by both \citet{2016ApJ...819L..17B} and \citet{2024MNRAS.530.2114G}. When using the updated migration torque values by \citet{2017MNRAS.471.4917J} for the \SK model, we find that the migration torque is always negative and thus the migrator  moves across the disc without being trapped. This result is in agreement with those by \citet{2024MNRAS.530.2114G}. Once the thermal torque from Eq.~(\ref{eq:Migrationtherm}) is added to the updated migration torque of Eq.~(\ref{eq:MigrationJimeneztot}), the bottom panel in Fig.~\ref{fig:migrationtorques} shows that we again obtain migration traps. %
In the \SK AGN disc, we find two migration traps for a $M=10^6 \,{\rm M}_\odot$ central BH and a $10 M_{\odot}$ migrator occurring at $r\approx 1.4 \times 10^3 R_{\rm s} = 1.4 \times 10^{-4}$ pc and $r\approx 6.8 \times 10^4 R_{\rm s} = 6.5 \times 10^{-3}$ pc. 

When considering the \TPO model, we obtain a larger number of migration traps, irrespective of the torque prescriptions adopted and the inclusion of the thermal torque contribution to the net torque. %
For the migration torque by \citet{2017MNRAS.471.4917J} (the top panel in Fig.~\ref{fig:migrationtorques}), we find migration traps form in the \TP disc when the gradient $\der \ln \Sigma_{\rm g} / \der \ln r$ discretely changes values, as can be seen in the lower panels of Fig.~\ref{fig:migrationtorques} at $r \approx 2.5 \times 10^3 R_{\rm s}$ and $r \approx  8.3 \times 10^3 R_{\rm s}$.
When both migration and thermal torques are considered, we find traps at $r\approx 2.5 \times 10^3 R_{\rm s}$, $r \approx 8.4 \times 10^3 R_{\rm s}$ and $r\approx 2.6 \times 10^6 R_{\rm s}$ 
for the \TPO model.

\section{Public implementation} 
\label{sec:code}

Our implementation of both the \SK and \TP models is released publicly in the \codename module for the Python programming language.

\codename is distributed under git version control at  
\begin{quote}
\href{https://github.com/DariaGangardt/pAGN}{github.com/DariaGangardt/pAGN} (code repository)
\end{quote}
The documentation is provided at  
 \begin{quote}
 \href{https://dariagangardt.github.io/pAGN}{dariagangardt.github.io/pAGN} (documentation)
\end{quote}
together with a set of minimal examples.

Our \codename module is available on the Python Package index. The code can be installed  with
 \begin{quote}
\texttt{pip install pagn}
\end{quote}
Packages \texttt{numpy}, \texttt{scipy}, and \texttt{matplotlib} are specified as dependencies. The package is imported with
 \begin{quote}
\texttt{import pagn}
\end{quote}
and contains two main classes for the \SK and \TP implementation, respectively:
 \begin{quote}
\texttt{pagn.SirkoAGN}\\
\phantom{x}\texttt{pagn.ThompsonAGN}
 \end{quote}
\noindent In addition, the code distributions include opacity tables by \citet{2003A&A...410..611S} and \citet{2005MNRAS.360..458B} as well as an interpolation routine. External opacity tables can also be provided by the user. %
The overall solution strategy follows what is presented in this paper as illustrated in the flowcharts of Figs.~\ref{fig:sirko_flowchart} and \ref{fig:thompson_flowchart}.

\section{Conclusions}
\label{sec:concl}

This work presents a critical re-analysis of  the AGN disc models by 
\SKO and \TPO. Our findings are implemented in the public \codename module for the Python programming language \citep{repo}. We presented the equations from the original papers and emphasized their solution strategy. Compared to the original model, our results consider updated opacity tables, relate some of the  input parameter (most notably the scaling of the outer accretion rate with the central BH mass for \TP case), validate AGN discs through limits on the accretion rate at the disc boundaries, and investigate the  limits of the thin-disc approximation. 
While the parameter exploration presented in this work provides valuable insights, there is room for further enhancement to fully explore the predictions of these models across the entire parameter space. An example of such research, \citet{2008ApJ...685..787B} presented an observation-motivated study of how the \TP input parameters affects the properties of AGN discs in Seyfert-like with a particular focus on the ``starburst'' disc regions with high star formation.

As a further example, in this paper we have applied our \codename code to the disc-migration problem, reproducing the analysis by \citet{2024MNRAS.530.2114G} with the more complex disc profiles by \SKO and \TPO. %
While we largely confirm previous findings for the \SK case, our \TP disc shows a large number of migration traps, with potential implications for the formation of hierarchical merging stellar-mass BH binaries detectable with current gravitational-wave detectors \citep{2021NatAs...5..749G}. This is an interesting avenue for future work.

The AGN disc models by \SKO and \TPO are widely used in the literature. We hope our full, public implementation of these approaches,  
together with the details of the underlying evolutionary equations, might facilitate further advances in this area while clarifying their underlying limitations. Both the \SK and \TP models can be applied to various problems and compared to newer AGN-disc modeling approaches. The goal of \codename is precisely that to aid further research in the growing field of AGN and gravitational-wave science.

\section*{Acknowledgements}

We thank Maria Paola Vaccaro, Andrea Derdzinski, Evgeni Grishin, Ari Laor, Hiromichi Tagawa, Shmuel Gilbaum, Sean McGee, Cressida Cleland %
for discussions.
D.Gan. and D.Ger. are supported by ERC Starting Grant No.~945155--GWmining, 
Cariplo Foundation Grant No.~2021-0555, MUR PRIN Grant No.~2022-Z9X4XS, 
and the ICSC National Research Centre funded by NextGenerationEU. 
A.A.T. acknowledges support from JSPS KAKENHI Grant No.~21K13914 and from the European Union’s Horizon 2020 and Horizon Europe research and innovation programs under the Marie Sk\l{}odowska-Curie grant agreements No.~847523 and No.~101103134.
D.Ger. is supported by MSCA Fellowships No.~101064542--StochRewind and No.~101149270--ProtoBH.
Computational work was performed at CINECA with allocations 
through INFN and Bicocca.%

\section*{Data Availability}
The data underlying this article are available at \cite{repo}. Additional data will be shared on reasonable request to the corresponding author.

\appendix
\section{Optically Thick Approximation in \TP} \label{sec:optthick}

For the \TP model, the case where the disc optically thick to its own infrared radiation can be approximated analytically. We assume that the gas is a constant fraction $f_\mathrm{g} \equiv \Sigma_\mathrm{g} / \Sigma_\mathrm{tot}$ 
of the total dynamical mass 
\begin{equation}
    \Sigma_\mathrm{tot} = \frac{\sigma^2}{\pi G r} \, .
\end{equation}
These assumption function best at large scales (i.e., $r \gg \Rout$) where the angular frequency is dominated by the velocity dispersion, so that $\Omega_\mathrm{T} \approx \sqrt{2} \sigma / r$. The mass density from Eq.~(\ref{eq:Thomrhoout}) reads
\begin{equation}
    \rho \approx \frac{\sqrt{2} \sigma^2}{\pi G Q_\mathrm{T} r^2} \, .
\end{equation}
 If $f_\mathrm{g}$ is constant, then the mid-scale height is given by
\begin{equation}
    \frac{h}{r} = \frac{f_\mathrm{g} Q_\mathrm{T}}{2^{3/2}} \, , 
\end{equation}
and the sound speed is
\begin{equation}
    \frac{c_\mathrm{s}}{\sigma} = \frac{f_\mathrm{g} Q_\mathrm{T}}{2} \, .
\end{equation}

At large values of $r$, the disc is mostly radiation-pressure dominated, so that $p_\mathrm{rad} = 4 \sigma_\mathrm{SB} T^4 / 3 c = \sigma_\mathrm{SB} \tauv \Teff^4 / c$. In the optically thick limit, the main contribution to $\Teff$ that of star formation
\begin{equation}
    \sigma_\mathrm{SB} \Teff^4 = \frac{1}{2} \epsilon_\mathrm{T} \dot{\Sigma}_* c^2 \, .
\end{equation}
Combining these equations, one can then find the temperature 
\begin{equation}
    T = \left( \frac{3c Q_\mathrm{T}}{2^{7/2} \pi G \sigma_\mathrm{SB}} \right)^{1/4} \left( \frac{f_\mathrm{g} \sigma^2}{r} \right)^{1/2} \, ,
\end{equation}
and the star formation rate
\begin{equation}
    \dot{\Sigma}_* = \frac{\sqrt{2}f_\mathrm{g} Q_\mathrm{T}}{\epsilon_\mathrm{T} \kappa c} \frac{\sigma^2}{r} \, .
\end{equation}

\bibliographystyle{mnras_tex_edited}
\bibliography{agnmodels}

\begin{thebibliography}{}
\makeatletter
\relax
\def\mn@urlcharsother{\let\do\@makeother \do\$\do\&\do\#\do\^\do\_\do\%\do\~}
\def\mn@doi{\begingroup\mn@urlcharsother \@ifnextchar [ {\mn@doi@}
  {\mn@doi@[]}}
\def\mn@doi@[#1]#2{\def\@tempa{#1}\ifx\@tempa\@empty \href
  {http://dx.doi.org/#2} {doi:#2}\else \href {http://dx.doi.org/#2} {#1}\fi
  \endgroup}
\def\mn@eprint#1#2{\mn@eprint@#1:#2::\@nil}
\def\mn@eprint@arXiv#1{\href {http://arxiv.org/abs/#1} {{arXiv:#1}}}
\def\mn@eprint@dblp#1{\href {http://dblp.uni-trier.de/rec/bibtex/#1.xml}
  {dblp:#1}}
\def\mn@eprint@#1:#2:#3:#4\@nil{\def\@tempa {#1}\def\@tempb {#2}\def\@tempc
  {#3}\ifx \@tempc \@empty \let \@tempc \@tempb \let \@tempb \@tempa \fi \ifx
  \@tempb \@empty \def\@tempb {arXiv}\fi \@ifundefined
  {mn@eprint@\@tempb}{\@tempb:\@tempc}{\expandafter \expandafter \csname
  mn@eprint@\@tempb\endcsname \expandafter{\@tempc}}}

\bibitem[\protect\citeauthoryear{{Alexander} \& {Ferguson}}{{Alexander} \&
  {Ferguson}}{1994}]{1994ApJ...437..879A}
{Alexander} D.~R.,  {Ferguson} J.~W.,  1994, \mn@doi [Astrophys. J.]
  {10.1086/175039}, \href
  {https://ui.adsabs.harvard.edu/abs/1994ApJ...437..879A} {437, 879}

\bibitem[\protect\citeauthoryear{{Armitage}}{{Armitage}}{2020}]{2020apfs.book.....A}
{Armitage} P.~J.,  2020, {Astrophysics of planet formation}.
Cambridge University Press

\bibitem[\protect\citeauthoryear{{Artymowicz}, {Lin}  \&
  {Wampler}}{{Artymowicz} et~al.}{1993}]{1993ApJ...409..592A}
{Artymowicz} P.,  {Lin} D.~N.~C.,   {Wampler} E.~J.,  1993, \mn@doi [Astrophys.
  J.] {10.1086/172690}, \href
  {https://ui.adsabs.harvard.edu/abs/1993ApJ...409..592A} {409, 592}

\bibitem[\protect\citeauthoryear{{Badnell}, {Bautista}, {Butler}, {Delahaye},
  {Mendoza}, {Palmeri}, {Zeippen}  \& {Seaton}}{{Badnell}
  et~al.}{2005}]{2005MNRAS.360..458B}
{Badnell} N.~R.,  {Bautista} M.~A.,  {Butler} K.,  {Delahaye} F.,  {Mendoza}
  C.,  {Palmeri} P.,  {Zeippen} C.~J.,   {Seaton} M.~J.,  2005, \mn@doi [Mon.
  Not. R. Astron. Soc.] {10.1111/j.1365-2966.2005.08991.x}, \href
  {https://ui.adsabs.harvard.edu/abs/2005MNRAS.360..458B} {360, 458}
  (\mn@eprint {arXiv} {astro-ph/0410744})

\bibitem[\protect\citeauthoryear{{Ballantyne}}{{Ballantyne}}{2008}]{2008ApJ...685..787B}
{Ballantyne} D.~R.,  2008, \mn@doi [Astrophys. J.] {10.1086/591048}, \href
  {https://ui.adsabs.harvard.edu/abs/2008ApJ...685..787B} {685, 787}
  (\mn@eprint {arXiv} {0806.2863})

\bibitem[\protect\citeauthoryear{{Barausse}, {Shankar}, {Bernardi}, {Dubois}
  \& {Sheth}}{{Barausse} et~al.}{2017}]{2017MNRAS.468.4782B}
{Barausse} E.,  {Shankar} F.,  {Bernardi} M.,  {Dubois} Y.,   {Sheth} R.~K.,
  2017, \mn@doi [Mon. Not. R. Astron. Soc.] {10.1093/mnras/stx799}, \href
  {https://ui.adsabs.harvard.edu/abs/2017MNRAS.468.4782B} {468, 4782}
  (\mn@eprint {arXiv} {1702.01762})

\bibitem[\protect\citeauthoryear{{Baskin} \& {Laor}}{{Baskin} \&
  {Laor}}{2018}]{2018MNRAS.474.1970B}
{Baskin} A.,  {Laor} A.,  2018, \mn@doi [\mnras] {10.1093/mnras/stx2850}, \href
  {https://ui.adsabs.harvard.edu/abs/2018MNRAS.474.1970B} {474, 1970}
  (\mn@eprint {arXiv} {1711.00025})

\bibitem[\protect\citeauthoryear{{Bellovary}, {Mac Low}, {McKernan}  \&
  {Ford}}{{Bellovary} et~al.}{2016}]{2016ApJ...819L..17B}
{Bellovary} J.~M.,  {Mac Low} M.-M.,  {McKernan} B.,   {Ford} K.~E.~S.,  2016,
  \mn@doi [Astrophys. J. Lett.] {10.3847/2041-8205/819/2/L17}, \href
  {https://ui.adsabs.harvard.edu/abs/2016ApJ...819L..17B} {819, L17}
  (\mn@eprint {arXiv} {1511.00005})

\bibitem[\protect\citeauthoryear{{Ben{\'\i}tez-Llambay}, {Masset},
  {Koenigsberger}  \& {Szul{\'a}gyi}}{{Ben{\'\i}tez-Llambay}
  et~al.}{2015}]{2015Natur.520...63B}
{Ben{\'\i}tez-Llambay} P.,  {Masset} F.,  {Koenigsberger} G.,   {Szul{\'a}gyi}
  J.,  2015, \mn@doi [Nature] {10.1038/nature14277}, \href
  {https://ui.adsabs.harvard.edu/abs/2015Natur.520...63B} {520, 63} (\mn@eprint
  {arXiv} {1510.01778})

\bibitem[\protect\citeauthoryear{{Bianchi}, {Mainieri}  \&
  {Padovani}}{{Bianchi} et~al.}{2022}]{2022hxga.book....4B}
{Bianchi} S.,  {Mainieri} V.,   {Padovani} P.,  2022, \mn@doi [Handbook of
  X-ray and Gamma-ray Astrophysics] {10.1007/978-981-16-4544-0_113-1}, \href
  {https://ui.adsabs.harvard.edu/abs/2022hxga.book....4B} {4}

\bibitem[\protect\citeauthoryear{{Burtscher} et~al.,}{{Burtscher}
  et~al.}{2013}]{2013A&A...558A.149B}
{Burtscher} L.,  et~al., 2013, \mn@doi [Astron. Astrophys.]
  {10.1051/0004-6361/201321890}, \href
  {https://ui.adsabs.harvard.edu/abs/2013A&A...558A.149B} {558, A149}
  (\mn@eprint {arXiv} {1307.2068})

\bibitem[\protect\citeauthoryear{{Cantiello}, {Jermyn}  \& {Lin}}{{Cantiello}
  et~al.}{2021}]{2021ApJ...910...94C}
{Cantiello} M.,  {Jermyn} A.~S.,   {Lin} D. N.~C.,  2021, \mn@doi [Astrophys.
  J.] {10.3847/1538-4357/abdf4f}, \href
  {https://ui.adsabs.harvard.edu/abs/2021ApJ...910...94C} {910, 94} (\mn@eprint
  {arXiv} {2009.03936})

\bibitem[\protect\citeauthoryear{{Derdzinski} \& {Mayer}}{{Derdzinski} \&
  {Mayer}}{2023}]{2023MNRAS.521.4522D}
{Derdzinski} A.,  {Mayer} L.,  2023, \mn@doi [\mnras] {10.1093/mnras/stad749},
  \href {https://ui.adsabs.harvard.edu/abs/2023MNRAS.521.4522D} {521, 4522}
  (\mn@eprint {arXiv} {2205.10382})

\bibitem[\protect\citeauthoryear{{Fabj}, {Nasim}, {Caban}, {Ford}, {McKernan}
  \& {Bellovary}}{{Fabj} et~al.}{2020}]{2020MNRAS.499.2608F}
{Fabj} G.,  {Nasim} S.~S.,  {Caban} F.,  {Ford} K.~E.~S.,  {McKernan} B.,
  {Bellovary} J.~M.,  2020, \mn@doi [Mon. Not. R. Astron. Soc.]
  {10.1093/mnras/staa3004}, \href
  {https://ui.adsabs.harvard.edu/abs/2020MNRAS.499.2608F} {499, 2608}
  (\mn@eprint {arXiv} {2006.11229})

\bibitem[\protect\citeauthoryear{{Falle}}{{Falle}}{1991}]{1991MNRAS.250..581F}
{Falle} S.~A.~E.~G.,  1991, \mn@doi [Mon. Not. R. Astron. Soc.]
  {10.1093/mnras/250.3.581}, \href
  {https://ui.adsabs.harvard.edu/abs/1991MNRAS.250..581F} {250, 581}

\bibitem[\protect\citeauthoryear{{Feldman} \& {Lin}}{{Feldman} \&
  {Lin}}{1973}]{1973StAM...52....1F}
{Feldman} S.~I.,  {Lin} C.~C.,  1973, Stud. Appl. Math., \href
  {https://ui.adsabs.harvard.edu/abs/1973StAM...52....1F} {52, 1}

\bibitem[\protect\citeauthoryear{{Gangardt} \& {Trani}}{{Gangardt} \&
  {Trani}}{2024}]{repo}
{Gangardt} D.,  {Trani} A.~A.,  2024,
  \href{https://doi.org/10.5281/zenodo.10723301}{doi.org/10.5281/zenodo.10723301},
  \\\href{https://github.com/DariaGangardt/pAGN}{github.com/DariaGangardt/pAGN}

\bibitem[\protect\citeauthoryear{{Garc{\'\i}a-Burillo}
  et~al.,}{{Garc{\'\i}a-Burillo} et~al.}{2019}]{2019A&A...632A..61G}
{Garc{\'\i}a-Burillo} S.,  et~al., 2019, \mn@doi [Astron. Astrophys.]
  {10.1051/0004-6361/201936606}, \href
  {https://ui.adsabs.harvard.edu/abs/2019A&A...632A..61G} {632, A61}
  (\mn@eprint {arXiv} {1909.00675})

\bibitem[\protect\citeauthoryear{{Gerosa} \& {Fishbach}}{{Gerosa} \&
  {Fishbach}}{2021}]{2021NatAs...5..749G}
{Gerosa} D.,  {Fishbach} M.,  2021, \mn@doi [Nat. Astron.]
  {10.1038/s41550-021-01398-w}, \href
  {https://ui.adsabs.harvard.edu/abs/2021NatAs...5..749G} {5, 749} (\mn@eprint
  {arXiv} {2105.03439})

\bibitem[\protect\citeauthoryear{{Gilbaum} \& {Stone}}{{Gilbaum} \&
  {Stone}}{2022}]{2022ApJ...928..191G}
{Gilbaum} S.,  {Stone} N.~C.,  2022, \mn@doi [Astrophys. J.]
  {10.3847/1538-4357/ac4ded}, \href
  {https://ui.adsabs.harvard.edu/abs/2022ApJ...928..191G} {928, 191}
  (\mn@eprint {arXiv} {2107.07519})

\bibitem[\protect\citeauthoryear{{Goldreich} \& {Tremaine}}{{Goldreich} \&
  {Tremaine}}{1979}]{1979ApJ...233..857G}
{Goldreich} P.,  {Tremaine} S.,  1979, \mn@doi [Astrophys. J.]
  {10.1086/157448}, \href
  {https://ui.adsabs.harvard.edu/abs/1979ApJ...233..857G} {233, 857}

\bibitem[\protect\citeauthoryear{{Goodman}}{{Goodman}}{2003}]{2003MNRAS.339..937G}
{Goodman} J.,  2003, \mn@doi [Mon. Not. R. Astron. Soc.]
  {10.1046/j.1365-8711.2003.06241.x}, \href
  {https://ui.adsabs.harvard.edu/abs/2003MNRAS.339..937G} {339, 937}
  (\mn@eprint {arXiv} {astro-ph/0201001})

\bibitem[\protect\citeauthoryear{{Grishin}, {Gilbaum}  \& {Stone}}{{Grishin}
  et~al.}{2024}]{2024MNRAS.530.2114G}
{Grishin} E.,  {Gilbaum} S.,   {Stone} N.~C.,  2024, \mn@doi [\mnras]
  {10.1093/mnras/stae828}, \href
  {https://ui.adsabs.harvard.edu/abs/2024MNRAS.530.2114G} {530, 2114}
  (\mn@eprint {arXiv} {2307.07546})

\bibitem[\protect\citeauthoryear{{Guilera}, {Miller Bertolami}, {Masset},
  {Cuadra}, {Venturini}  \& {Ronco}}{{Guilera}
  et~al.}{2021}]{2021MNRAS.507.3638G}
{Guilera} O.~M.,  {Miller Bertolami} M.~M.,  {Masset} F.,  {Cuadra} J.,
  {Venturini} J.,   {Ronco} M.~P.,  2021, \mn@doi [\mnras]
  {10.1093/mnras/stab2371}, \href
  {https://ui.adsabs.harvard.edu/abs/2021MNRAS.507.3638G} {507, 3638}
  (\mn@eprint {arXiv} {2108.04880})

\bibitem[\protect\citeauthoryear{{G{\"u}ltekin} et~al.,}{{G{\"u}ltekin}
  et~al.}{2009}]{2009ApJ...698..198G}
{G{\"u}ltekin} K.,  et~al., 2009, \mn@doi [Astrophys. J.]
  {10.1088/0004-637X/698/1/198}, \href
  {https://ui.adsabs.harvard.edu/abs/2009ApJ...698..198G} {698, 198}
  (\mn@eprint {arXiv} {0903.4897})

\bibitem[\protect\citeauthoryear{{Guo}, {Li}, {Zhang}, {Ho}  \& {Wang}}{{Guo}
  et~al.}{2022a}]{2022ApJ...929...19G}
{Guo} W.-J.,  {Li} Y.-R.,  {Zhang} Z.-X.,  {Ho} L.~C.,   {Wang} J.-M.,  2022a,
  \mn@doi [Astrophys. J.] {10.3847/1538-4357/ac4e84}, \href
  {https://ui.adsabs.harvard.edu/abs/2022ApJ...929...19G} {929, 19} (\mn@eprint
  {arXiv} {2201.08533})

\bibitem[\protect\citeauthoryear{{Guo}, {Barth}  \& {Wang}}{{Guo}
  et~al.}{2022b}]{2022ApJ...940...20G}
{Guo} H.,  {Barth} A.~J.,   {Wang} S.,  2022b, \mn@doi [Astrophys. J.]
  {10.3847/1538-4357/ac96ec}, \href
  {https://ui.adsabs.harvard.edu/abs/2022ApJ...940...20G} {940, 20} (\mn@eprint
  {arXiv} {2207.06432})

\bibitem[\protect\citeauthoryear{{Haiman}, {Kocsis}  \& {Menou}}{{Haiman}
  et~al.}{2009}]{2009ApJ...700.1952H}
{Haiman} Z.,  {Kocsis} B.,   {Menou} K.,  2009, \mn@doi [Astrophys. J.]
  {10.1088/0004-637X/700/2/1952}, \href
  {https://ui.adsabs.harvard.edu/abs/2009ApJ...700.1952H} {700, 1952}
  (\mn@eprint {arXiv} {0904.1383})

\bibitem[\protect\citeauthoryear{{Hawley}, {Smarr}  \& {Wilson}}{{Hawley}
  et~al.}{1984}]{1984ApJ...277..296H}
{Hawley} J.~F.,  {Smarr} L.~L.,   {Wilson} J.~R.,  1984, \mn@doi [Astrophys.
  J.] {10.1086/161696}, \href
  {https://ui.adsabs.harvard.edu/abs/1984ApJ...277..296H} {277, 296}

\bibitem[\protect\citeauthoryear{{Heckman}, {Kauffmann}, {Brinchmann},
  {Charlot}, {Tremonti}  \& {White}}{{Heckman}
  et~al.}{2004}]{2004ApJ...613..109H}
{Heckman} T.~M.,  {Kauffmann} G.,  {Brinchmann} J.,  {Charlot} S.,  {Tremonti}
  C.,   {White} S. D.~M.,  2004, \mn@doi [Astrophys. J.] {10.1086/422872},
  \href {https://ui.adsabs.harvard.edu/abs/2004ApJ...613..109H} {613, 109}
  (\mn@eprint {arXiv} {astro-ph/0406218})

\bibitem[\protect\citeauthoryear{{Hickox} \& {Alexander}}{{Hickox} \&
  {Alexander}}{2018}]{2018ARA&A..56..625H}
{Hickox} R.~C.,  {Alexander} D.~M.,  2018, \mn@doi [Annu. Rev. Astron.
  Astrophys.] {10.1146/annurev-astro-081817-051803}, \href
  {https://ui.adsabs.harvard.edu/abs/2018ARA&A..56..625H} {56, 625} (\mn@eprint
  {arXiv} {1806.04680})

\bibitem[\protect\citeauthoryear{{Hopkins} et~al.,}{{Hopkins}
  et~al.}{2024a}]{2024OJAp....7E..19H}
{Hopkins} P.~F.,  et~al., 2024a, \mn@doi [The Open Journal of Astrophysics]
  {10.21105/astro.2310.04506}, \href
  {https://ui.adsabs.harvard.edu/abs/2024OJAp....7E..19H} {7, 19} (\mn@eprint
  {arXiv} {2310.04506})

\bibitem[\protect\citeauthoryear{{Hopkins} et~al.,}{{Hopkins}
  et~al.}{2024b}]{2024OJAp....7E..20H}
{Hopkins} P.~F.,  et~al., 2024b, \mn@doi [The Open Journal of Astrophysics]
  {10.21105/astro.2310.04507}, \href
  {https://ui.adsabs.harvard.edu/abs/2024OJAp....7E..20H} {7, 20} (\mn@eprint
  {arXiv} {2310.04507})

\bibitem[\protect\citeauthoryear{{Hu{\v{s}}ko} \& {Lacey}}{{Hu{\v{s}}ko} \&
  {Lacey}}{2023}]{2023MNRAS.520.5090H}
{Hu{\v{s}}ko} F.,  {Lacey} C.~G.,  2023, \mn@doi [Mon. Not. R. Astron. Soc.]
  {10.1093/mnras/stad450}, \href
  {https://ui.adsabs.harvard.edu/abs/2023MNRAS.520.5090H} {520, 5090}
  (\mn@eprint {arXiv} {2205.08884})

\bibitem[\protect\citeauthoryear{{Iglesias} \& {Rogers}}{{Iglesias} \&
  {Rogers}}{1996}]{1996ApJ...464..943I}
{Iglesias} C.~A.,  {Rogers} F.~J.,  1996, \mn@doi [Astrophys. J.]
  {10.1086/177381}, \href
  {https://ui.adsabs.harvard.edu/abs/1996ApJ...464..943I} {464, 943}

\bibitem[\protect\citeauthoryear{{Jaffe} et~al.,}{{Jaffe}
  et~al.}{2004}]{2004Natur.429...47J}
{Jaffe} W.,  et~al., 2004, \mn@doi [Nature] {10.1038/nature02531}, \href
  {https://ui.adsabs.harvard.edu/abs/2004Natur.429...47J} {429, 47}

\bibitem[\protect\citeauthoryear{{Jha}, {Joshi}, {Chand}, {Wu}, {Ho}, {Rastogi}
   \& {Ma}}{{Jha} et~al.}{2022}]{2022MNRAS.511.3005J}
{Jha} V.~K.,  {Joshi} R.,  {Chand} H.,  {Wu} X.-B.,  {Ho} L.~C.,  {Rastogi} S.,
    {Ma} Q.,  2022, \mn@doi [Mon. Not. R. Astron. Soc.]
  {10.1093/mnras/stac109}, \href
  {https://ui.adsabs.harvard.edu/abs/2022MNRAS.511.3005J} {511, 3005}
  (\mn@eprint {arXiv} {2109.05036})

\bibitem[\protect\citeauthoryear{{Jiang}, {Davis}  \& {Stone}}{{Jiang}
  et~al.}{2016}]{2016ApJ...827...10J}
{Jiang} Y.-F.,  {Davis} S.~W.,   {Stone} J.~M.,  2016, \mn@doi [Astrophys. J.]
  {10.3847/0004-637X/827/1/10}, \href
  {https://ui.adsabs.harvard.edu/abs/2016ApJ...827...10J} {827, 10} (\mn@eprint
  {arXiv} {1601.06836})

\bibitem[\protect\citeauthoryear{{Jim{\'e}nez} \& {Masset}}{{Jim{\'e}nez} \&
  {Masset}}{2017}]{2017MNRAS.471.4917J}
{Jim{\'e}nez} M.~A.,  {Masset} F.~S.,  2017, \mn@doi [Mon. Not. R. Astron.
  Soc.] {10.1093/mnras/stx1946}, \href
  {https://ui.adsabs.harvard.edu/abs/2017MNRAS.471.4917J} {471, 4917}
  (\mn@eprint {arXiv} {1707.08988})

\bibitem[\protect\citeauthoryear{{Kley} \& {Crida}}{{Kley} \&
  {Crida}}{2008}]{2008A&A...487L...9K}
{Kley} W.,  {Crida} A.,  2008, \mn@doi [Astron. Astrophys.]
  {10.1051/0004-6361:200810033}, \href
  {https://ui.adsabs.harvard.edu/abs/2008A&A...487L...9K} {487, L9} (\mn@eprint
  {arXiv} {0806.2990})

\bibitem[\protect\citeauthoryear{{Kollmeier} et~al.,}{{Kollmeier}
  et~al.}{2006}]{2006ApJ...648..128K}
{Kollmeier} J.~A.,  et~al., 2006, \mn@doi [\apj] {10.1086/505646}, \href
  {https://ui.adsabs.harvard.edu/abs/2006ApJ...648..128K} {648, 128}
  (\mn@eprint {arXiv} {astro-ph/0508657})

\bibitem[\protect\citeauthoryear{{Kong} \& {Ho}}{{Kong} \&
  {Ho}}{2018}]{2018ApJ...859..116K}
{Kong} M.,  {Ho} L.~C.,  2018, \mn@doi [Astrophys. J.]
  {10.3847/1538-4357/aabe2a}, \href
  {https://ui.adsabs.harvard.edu/abs/2018ApJ...859..116K} {859, 116}
  (\mn@eprint {arXiv} {1804.09852})

\bibitem[\protect\citeauthoryear{{Korycansky} \& {Pollack}}{{Korycansky} \&
  {Pollack}}{1993}]{1993Icar..102..150K}
{Korycansky} D.~G.,  {Pollack} J.~B.,  1993, \mn@doi [Icarus]
  {10.1006/icar.1993.1039}, \href
  {https://ui.adsabs.harvard.edu/abs/1993Icar..102..150K} {102, 150}

\bibitem[\protect\citeauthoryear{{Lega}, {Crida}, {Bitsch}  \&
  {Morbidelli}}{{Lega} et~al.}{2014}]{2014MNRAS.440..683L}
{Lega} E.,  {Crida} A.,  {Bitsch} B.,   {Morbidelli} A.,  2014, \mn@doi [Mon.
  Not. R. Astron. Soc.] {10.1093/mnras/stu304}, \href
  {https://ui.adsabs.harvard.edu/abs/2014MNRAS.440..683L} {440, 683}
  (\mn@eprint {arXiv} {1402.2834})

\bibitem[\protect\citeauthoryear{{Levin}}{{Levin}}{2003}]{2003astro.ph..7084L}
{Levin} Y.,  2003, \mn@doi [arXiv e-prints] {10.48550/arXiv.astro-ph/0307084},
  \href {https://ui.adsabs.harvard.edu/abs/2003astro.ph..7084L} {pp
  astro--ph/0307084} (\mn@eprint {arXiv} {astro-ph/0307084})

\bibitem[\protect\citeauthoryear{{Lin} \& {Papaloizou}}{{Lin} \&
  {Papaloizou}}{1979}]{1979MNRAS.186..799L}
{Lin} D.~N.~C.,  {Papaloizou} J.,  1979, \mn@doi [Mon. Not. R. Astron. Soc.]
  {10.1093/mnras/186.4.799}, \href
  {https://ui.adsabs.harvard.edu/abs/1979MNRAS.186..799L} {186, 799}

\bibitem[\protect\citeauthoryear{{Lyra}, {Paardekooper}  \& {Mac Low}}{{Lyra}
  et~al.}{2010}]{2010ApJ...715L..68L}
{Lyra} W.,  {Paardekooper} S.-J.,   {Mac Low} M.-M.,  2010, \mn@doi [Astrophys.
  J. Lett.] {10.1088/2041-8205/715/2/L68}, \href
  {https://ui.adsabs.harvard.edu/abs/2010ApJ...715L..68L} {715, L68}
  (\mn@eprint {arXiv} {1003.0925})

\bibitem[\protect\citeauthoryear{{Masset}}{{Masset}}{2017}]{2017MNRAS.472.4204M}
{Masset} F.~S.,  2017, \mn@doi [Mon. Not. R. Astron. Soc.]
  {10.1093/mnras/stx2271}, \href
  {https://ui.adsabs.harvard.edu/abs/2017MNRAS.472.4204M} {472, 4204}
  (\mn@eprint {arXiv} {1708.09807})

\bibitem[\protect\citeauthoryear{{Masset} \& {Casoli}}{{Masset} \&
  {Casoli}}{2010}]{2010ApJ...723.1393M}
{Masset} F.~S.,  {Casoli} J.,  2010, \mn@doi [Astrophys. J.]
  {10.1088/0004-637X/723/2/1393}, \href
  {https://ui.adsabs.harvard.edu/abs/2010ApJ...723.1393M} {723, 1393}
  (\mn@eprint {arXiv} {1009.1913})

\bibitem[\protect\citeauthoryear{{McKernan}, {Ford}, {Lyra}, {Perets}, {Winter}
   \& {Yaqoob}}{{McKernan} et~al.}{2011}]{2011MNRAS.417L.103M}
{McKernan} B.,  {Ford} K.~E.~S.,  {Lyra} W.,  {Perets} H.~B.,  {Winter} L.~M.,
   {Yaqoob} T.,  2011, \mn@doi [Mon. Not. R. Astron. Soc.]
  {10.1111/j.1745-3933.2011.01132.x}, \href
  {https://ui.adsabs.harvard.edu/abs/2011MNRAS.417L.103M} {417, L103}
  (\mn@eprint {arXiv} {1108.1787})

\bibitem[\protect\citeauthoryear{{McKernan}, {Ford}, {Lyra}  \&
  {Perets}}{{McKernan} et~al.}{2012}]{2012MNRAS.425..460M}
{McKernan} B.,  {Ford} K.~E.~S.,  {Lyra} W.,   {Perets} H.~B.,  2012, \mn@doi
  [Mon. Not. R. Astron. Soc.] {10.1111/j.1365-2966.2012.21486.x}, \href
  {https://ui.adsabs.harvard.edu/abs/2012MNRAS.425..460M} {425, 460}
  (\mn@eprint {arXiv} {1206.2309})

\bibitem[\protect\citeauthoryear{{Menci}, {Fiore}, {Shankar}, {Zanisi}  \&
  {Feruglio}}{{Menci} et~al.}{2023}]{2023A&A...674A.181M}
{Menci} N.,  {Fiore} F.,  {Shankar} F.,  {Zanisi} L.,   {Feruglio} C.,  2023,
  \mn@doi [Astron. Astrophys.] {10.1051/0004-6361/202244574}, \href
  {https://ui.adsabs.harvard.edu/abs/2023A&A...674A.181M} {674, A181}
  (\mn@eprint {arXiv} {2304.08273})

\bibitem[\protect\citeauthoryear{{Murray}, {Quataert}  \& {Thompson}}{{Murray}
  et~al.}{2005}]{2005ApJ...618..569M}
{Murray} N.,  {Quataert} E.,   {Thompson} T.~A.,  2005, \mn@doi [Astrophys. J.]
  {10.1086/426067}, \href
  {https://ui.adsabs.harvard.edu/abs/2005ApJ...618..569M} {618, 569}
  (\mn@eprint {arXiv} {astro-ph/0406070})

\bibitem[\protect\citeauthoryear{{Nelson}, {Papaloizou}, {Masset}  \&
  {Kley}}{{Nelson} et~al.}{2000}]{2000MNRAS.318...18N}
{Nelson} R.~P.,  {Papaloizou} J. C.~B.,  {Masset} F.,   {Kley} W.,  2000,
  \mn@doi [Mon. Not. R. Astron. Soc.] {10.1046/j.1365-8711.2000.03605.x}, \href
  {https://ui.adsabs.harvard.edu/abs/2000MNRAS.318...18N} {318, 18} (\mn@eprint
  {arXiv} {astro-ph/9909486})

\bibitem[\protect\citeauthoryear{{Netzer}}{{Netzer}}{2015}]{2015ARA&A..53..365N}
{Netzer} H.,  2015, \mn@doi [Annu. Rev. Astron. Astrophys.]
  {10.1146/annurev-astro-082214-122302}, \href
  {https://ui.adsabs.harvard.edu/abs/2015ARA&A..53..365N} {53, 365} (\mn@eprint
  {arXiv} {1505.00811})

\bibitem[\protect\citeauthoryear{{Paardekooper} \& {Mellema}}{{Paardekooper} \&
  {Mellema}}{2006}]{2006A&A...459L..17P}
{Paardekooper} S.~J.,  {Mellema} G.,  2006, \mn@doi [Astron. Astrophys.]
  {10.1051/0004-6361:20066304}, \href
  {https://ui.adsabs.harvard.edu/abs/2006A&A...459L..17P} {459, L17}
  (\mn@eprint {arXiv} {astro-ph/0608658})

\bibitem[\protect\citeauthoryear{{Paardekooper}, {Baruteau}, {Crida}  \&
  {Kley}}{{Paardekooper} et~al.}{2010}]{2010MNRAS.401.1950P}
{Paardekooper} S.~J.,  {Baruteau} C.,  {Crida} A.,   {Kley} W.,  2010, \mn@doi
  [Mon. Not. R. Astron. Soc.] {10.1111/j.1365-2966.2009.15782.x}, \href
  {https://ui.adsabs.harvard.edu/abs/2010MNRAS.401.1950P} {401, 1950}
  (\mn@eprint {arXiv} {0909.4552})

\bibitem[\protect\citeauthoryear{{Padovani} et~al.,}{{Padovani}
  et~al.}{2017}]{2017A&ARv..25....2P}
{Padovani} P.,  et~al., 2017, \mn@doi [Astron. Astrophys. Rev.]
  {10.1007/s00159-017-0102-9}, \href
  {https://ui.adsabs.harvard.edu/abs/2017A&ARv..25....2P} {25, 2} (\mn@eprint
  {arXiv} {1707.07134})

\bibitem[\protect\citeauthoryear{{Pringle}}{{Pringle}}{1981}]{1981ARA&A..19..137P}
{Pringle} J.~E.,  1981, \mn@doi [Annu. Rev. Astron. Astrophys.]
  {10.1146/annurev.aa.19.090181.001033}, \href
  {https://ui.adsabs.harvard.edu/abs/1981ARA&A..19..137P} {19, 137}

\bibitem[\protect\citeauthoryear{{Sajina}, {Lacy}  \& {Pope}}{{Sajina}
  et~al.}{2022}]{2022Univ....8..356S}
{Sajina} A.,  {Lacy} M.,   {Pope} A.,  2022, \mn@doi [Universe]
  {10.3390/universe8070356}, \href
  {https://ui.adsabs.harvard.edu/abs/2022Univ....8..356S} {8, 356} (\mn@eprint
  {arXiv} {2210.02307})

\bibitem[\protect\citeauthoryear{{Salpeter}}{{Salpeter}}{1964}]{1964ApJ...140..796S}
{Salpeter} E.~E.,  1964, \mn@doi [Astrophys. J.] {10.1086/147973}, \href
  {https://ui.adsabs.harvard.edu/abs/1964ApJ...140..796S} {140, 796}

\bibitem[\protect\citeauthoryear{{Santini}, {Gerosa}, {Cotesta}  \&
  {Berti}}{{Santini} et~al.}{2023}]{2023PhRvD.108h3033S}
{Santini} A.,  {Gerosa} D.,  {Cotesta} R.,   {Berti} E.,  2023, \mn@doi [Phys.
  Rev. D] {10.1103/PhysRevD.108.083033}, \href
  {https://ui.adsabs.harvard.edu/abs/2023PhRvD.108h3033S} {108, 083033}
  (\mn@eprint {arXiv} {2308.12998})

\bibitem[\protect\citeauthoryear{{Schartmann}, {Meisenheimer}, {Camenzind},
  {Wolf}  \& {Henning}}{{Schartmann} et~al.}{2005}]{2005A&A...437..861S}
{Schartmann} M.,  {Meisenheimer} K.,  {Camenzind} M.,  {Wolf} S.,   {Henning}
  T.,  2005, \mn@doi [Astron. Astrophys.] {10.1051/0004-6361:20042363}, \href
  {https://ui.adsabs.harvard.edu/abs/2005A&A...437..861S} {437, 861}
  (\mn@eprint {arXiv} {astro-ph/0504105})

\bibitem[\protect\citeauthoryear{{Schartmann}, {Meisenheimer}, {Camenzind},
  {Wolf}, {Tristram}  \& {Henning}}{{Schartmann}
  et~al.}{2008}]{2008A&A...482...67S}
{Schartmann} M.,  {Meisenheimer} K.,  {Camenzind} M.,  {Wolf} S.,  {Tristram}
  K.~R.~W.,   {Henning} T.,  2008, \mn@doi [Astron. Astrophys.]
  {10.1051/0004-6361:20078907}, \href
  {https://ui.adsabs.harvard.edu/abs/2008A&A...482...67S} {482, 67} (\mn@eprint
  {arXiv} {0802.2604})

\bibitem[\protect\citeauthoryear{{Secunda}, {Bellovary}, {Mac Low}, {Ford},
  {McKernan}, {Leigh}, {Lyra}  \& {S{\'a}ndor}}{{Secunda}
  et~al.}{2019}]{2019ApJ...878...85S}
{Secunda} A.,  {Bellovary} J.,  {Mac Low} M.-M.,  {Ford} K.~E.~S.,  {McKernan}
  B.,  {Leigh} N. W.~C.,  {Lyra} W.,   {S{\'a}ndor} Z.,  2019, \mn@doi
  [Astrophys. J.] {10.3847/1538-4357/ab20ca}, \href
  {https://ui.adsabs.harvard.edu/abs/2019ApJ...878...85S} {878, 85} (\mn@eprint
  {arXiv} {1807.02859})

\bibitem[\protect\citeauthoryear{{Semenov}, {Henning}, {Helling}, {Ilgner}  \&
  {Sedlmayr}}{{Semenov} et~al.}{2003}]{2003A&A...410..611S}
{Semenov} D.,  {Henning} T.,  {Helling} C.,  {Ilgner} M.,   {Sedlmayr} E.,
  2003, \mn@doi [Astron. Astrophys.] {10.1051/0004-6361:20031279}, \href
  {https://ui.adsabs.harvard.edu/abs/2003A&A...410..611S} {410, 611}
  (\mn@eprint {arXiv} {astro-ph/0308344})

\bibitem[\protect\citeauthoryear{{Shakura} \& {Sunyaev}}{{Shakura} \&
  {Sunyaev}}{1973}]{1973A&A....24..337S}
{Shakura} N.~I.,  {Sunyaev} R.~A.,  1973, Astron. Astrophys., \href
  {https://ui.adsabs.harvard.edu/abs/1973A&A....24..337S} {24, 337}

\bibitem[\protect\citeauthoryear{{Shankar}, {Bernardi}  \& {Sheth}}{{Shankar}
  et~al.}{2017}]{2017MNRAS.466.4029S}
{Shankar} F.,  {Bernardi} M.,   {Sheth} R.~K.,  2017, \mn@doi [Mon. Not. R.
  Astron. Soc.] {10.1093/mnras/stw3368}, \href
  {https://ui.adsabs.harvard.edu/abs/2017MNRAS.466.4029S} {466, 4029}
  (\mn@eprint {arXiv} {1701.01732})

\bibitem[\protect\citeauthoryear{{Sirko} \& {Goodman}}{{Sirko} \&
  {Goodman}}{2003}]{2003MNRAS.341..501S}
{Sirko} E.,  {Goodman} J.,  2003, \mn@doi [Mon. Not. R. Astron. Soc.]
  {10.1046/j.1365-8711.2003.06431.x}, \href
  {https://ui.adsabs.harvard.edu/abs/2003MNRAS.341..501S} {341, 501}
  (\mn@eprint {arXiv} {astro-ph/0209469})

\bibitem[\protect\citeauthoryear{{Stone}, {Metzger}  \& {Haiman}}{{Stone}
  et~al.}{2017}]{2017MNRAS.464..946S}
{Stone} N.~C.,  {Metzger} B.~D.,   {Haiman} Z.,  2017, \mn@doi [Mon. Not. R.
  Astron. Soc.] {10.1093/mnras/stw2260}, \href
  {https://ui.adsabs.harvard.edu/abs/2017MNRAS.464..946S} {464, 946}
  (\mn@eprint {arXiv} {1602.04226})

\bibitem[\protect\citeauthoryear{{Suh}, {Hasinger}, {Steinhardt}, {Silverman}
  \& {Schramm}}{{Suh} et~al.}{2015}]{2015ApJ...815..129S}
{Suh} H.,  {Hasinger} G.,  {Steinhardt} C.,  {Silverman} J.~D.,   {Schramm} M.,
   2015, \mn@doi [Astrophys. J.] {10.1088/0004-637X/815/2/129}, \href
  {https://ui.adsabs.harvard.edu/abs/2015ApJ...815..129S} {815, 129}
  (\mn@eprint {arXiv} {1511.01092})

\bibitem[\protect\citeauthoryear{{Syer}, {Clarke}  \& {Rees}}{{Syer}
  et~al.}{1991}]{1991MNRAS.250..505S}
{Syer} D.,  {Clarke} C.~J.,   {Rees} M.~J.,  1991, \mn@doi [Mon. Not. R.
  Astron. Soc.] {10.1093/mnras/250.3.505}, \href
  {https://ui.adsabs.harvard.edu/abs/1991MNRAS.250..505S} {250, 505}

\bibitem[\protect\citeauthoryear{{Tagawa}, {Haiman}  \& {Kocsis}}{{Tagawa}
  et~al.}{2020}]{2020ApJ...898...25T}
{Tagawa} H.,  {Haiman} Z.,   {Kocsis} B.,  2020, \mn@doi [Astrophys. J.]
  {10.3847/1538-4357/ab9b8c}, \href
  {https://ui.adsabs.harvard.edu/abs/2020ApJ...898...25T} {898, 25} (\mn@eprint
  {arXiv} {1912.08218})

\bibitem[\protect\citeauthoryear{{Tanaka}, {Takeuchi}  \& {Ward}}{{Tanaka}
  et~al.}{2002}]{2002ApJ...565.1257T}
{Tanaka} H.,  {Takeuchi} T.,   {Ward} W.~R.,  2002, \mn@doi [Astrophys. J.]
  {10.1086/324713}, \href
  {https://ui.adsabs.harvard.edu/abs/2002ApJ...565.1257T} {565, 1257}

\bibitem[\protect\citeauthoryear{{Thompson}, {Quataert}  \&
  {Murray}}{{Thompson} et~al.}{2005}]{2005ApJ...630..167T}
{Thompson} T.~A.,  {Quataert} E.,   {Murray} N.,  2005, \mn@doi [Astrophys. J.]
  {10.1086/431923}, \href
  {https://ui.adsabs.harvard.edu/abs/2005ApJ...630..167T} {630, 167}
  (\mn@eprint {arXiv} {astro-ph/0503027})

\bibitem[\protect\citeauthoryear{{Toomre}}{{Toomre}}{1964}]{1964ApJ...139.1217T}
{Toomre} A.,  1964, \mn@doi [Astrophys. J.] {10.1086/147861}, \href
  {https://ui.adsabs.harvard.edu/abs/1964ApJ...139.1217T} {139, 1217}

\bibitem[\protect\citeauthoryear{{Trani}, {Quaini}  \& {Colpi}}{{Trani}
  et~al.}{2024}]{2024A&A...683A.135T}
{Trani} A.~A.,  {Quaini} S.,   {Colpi} M.,  2024, \mn@doi [\aap]
  {10.1051/0004-6361/202347920}, \href
  {https://ui.adsabs.harvard.edu/abs/2024A&A...683A.135T} {683, A135}
  (\mn@eprint {arXiv} {2312.13281})

\bibitem[\protect\citeauthoryear{{Tremaine} et~al.,}{{Tremaine}
  et~al.}{2002}]{2002ApJ...574..740T}
{Tremaine} S.,  et~al., 2002, \mn@doi [Astrophys. J.] {10.1086/341002}, \href
  {https://ui.adsabs.harvard.edu/abs/2002ApJ...574..740T} {574, 740}
  (\mn@eprint {arXiv} {astro-ph/0203468})

\bibitem[\protect\citeauthoryear{{Vaccaro}, {Mapelli}, {P{\'e}rigois},
  {Barone}, {Artale}, {Dall'Amico}, {Iorio}  \& {Torniamenti}}{{Vaccaro}
  et~al.}{2023}]{2023arXiv231118548V}
{Vaccaro} M.~P.,  {Mapelli} M.,  {P{\'e}rigois} C.,  {Barone} D.,  {Artale}
  M.~C.,  {Dall'Amico} M.,  {Iorio} G.,   {Torniamenti} S.,  2023, \href
  {https://ui.adsabs.harvard.edu/abs/2023arXiv231118548V} {} (\mn@eprint
  {arXiv} {2311.18548})

\bibitem[\protect\citeauthoryear{{Velasco Romero} \& {Masset}}{{Velasco Romero}
  \& {Masset}}{2020}]{2020MNRAS.495.2063V}
{Velasco Romero} D.~A.,  {Masset} F.~S.,  2020, \mn@doi [\mnras]
  {10.1093/mnras/staa1215}, \href
  {https://ui.adsabs.harvard.edu/abs/2020MNRAS.495.2063V} {495, 2063}
  (\mn@eprint {arXiv} {2004.13422})

\bibitem[\protect\citeauthoryear{{Wada}, {Papadopoulos}  \& {Spaans}}{{Wada}
  et~al.}{2009}]{2009ApJ...702...63W}
{Wada} K.,  {Papadopoulos} P.~P.,   {Spaans} M.,  2009, \mn@doi [Astrophys. J.]
  {10.1088/0004-637X/702/1/63}, \href
  {https://ui.adsabs.harvard.edu/abs/2009ApJ...702...63W} {702, 63} (\mn@eprint
  {arXiv} {0906.5444})

\bibitem[\protect\citeauthoryear{{Ward}}{{Ward}}{1997}]{1997Icar..126..261W}
{Ward} W.~R.,  1997, \mn@doi [Icarus] {10.1006/icar.1996.5647}, \href
  {https://ui.adsabs.harvard.edu/abs/1997Icar..126..261W} {126, 261}

\bibitem[\protect\citeauthoryear{{Whitehead}, {Rowan}, {Boekholt}  \&
  {Kocsis}}{{Whitehead} et~al.}{2023}]{2023arXiv230911561W}
{Whitehead} H.,  {Rowan} C.,  {Boekholt} T.,   {Kocsis} B.,  2023, \href
  {https://ui.adsabs.harvard.edu/abs/2023arXiv230911561W} {} (\mn@eprint
  {arXiv} {2309.11561})

\bibitem[\protect\citeauthoryear{{Wilson}}{{Wilson}}{1972}]{1972ApJ...173..431W}
{Wilson} J.~R.,  1972, \mn@doi [Astrophys. J.] {10.1086/151434}, \href
  {https://ui.adsabs.harvard.edu/abs/1972ApJ...173..431W} {173, 431}

\bibitem[\protect\citeauthoryear{{Yang} et~al.,}{{Yang}
  et~al.}{2019}]{2019PhRvL.123r1101Y}
{Yang} Y.,  et~al., 2019, \mn@doi [Phys. Rev. Lett.]
  {10.1103/PhysRevLett.123.181101}, \href
  {https://ui.adsabs.harvard.edu/abs/2019PhRvL.123r1101Y} {123, 181101}
  (\mn@eprint {arXiv} {1906.09281})

\bibitem[\protect\citeauthoryear{{Zel'dovich}}{{Zel'dovich}}{1964}]{1964SPhD....9..195Z}
{Zel'dovich} Y.~B.,  1964, Sov. Phys. Dokl., \href
  {https://ui.adsabs.harvard.edu/abs/1964SPhD....9..195Z} {9, 195}

\makeatother
\end{thebibliography}

\end{document}